\documentclass[twocolumn, twocolappendix]{aastex631}

\usepackage{threeparttable}

\newcommand{\teff}{$T_{\rm eff}$}

\newcommand{\logg}{$\log g$}

\newcommand{\feh}{$\rm{[Fe/H]}$}

\newcommand{\vmic}{$v_{t}$}

\newcommand{\qq}{$\mathrm{q}^{2}$}

\newcommand{\sm}{$\rm{M_{\odot}}$}

\newcommand{\be}{\begin{equation}}
\newcommand{\ee}{\end{equation}}
\newcommand{\ben}{\begin{eqnarray}}
\newcommand{\een}{\end{eqnarray}}
\newcommand{\bfg}{\begin{figure}}
\newcommand{\efg}{\end{figure}}

\newcommand{\li}{$A$(Li)}
\newcommand{\linlte}{$A$(Li)$_{\rm{NLTE}}$}

\newcommand{\g}{$G$}

\newcommand{\Tc}{$T_{\rm{C}}$}

\shorttitle{Possible evidence of planet engulfment in TOI-1173 A/B}
\shortauthors{Yana Galarza et al.}

\begin{document}

\title{Detailed abundances of the planet-hosting TOI-1173 A/B system: \\
Possible evidence of planet engulfment in a very wide binary}

\correspondingauthor{J. Yana Galarza}
\email{jyanagalarza@carnegiescience.edu}

\author[0000-0001-9261-8366]{Jhon Yana Galarza}
\altaffiliation{Carnegie Fellow}
\affiliation{The Observatories of the Carnegie Institution for Science, 813 Santa Barbara Street, Pasadena, CA 91101, USA}

\author[0000-0001-6533-6179]{Henrique Reggiani}
\affiliation{Gemini South, Gemini Observatory, NSF's NOIRLab, Casilla 603, La Serena, Chile}

\author[0000-0003-2059-470X]{Thiago Ferreira}
\affiliation{Department of Astronomy, Yale University, 219 Prospect St., New Haven, CT 06511, USA}

\author[0000-0002-1387-2954]{Diego Lorenzo-Oliveira}
\affiliation{Laborat\'orio Nacional de Astrof\'isica, Rua Estados Unidos 154, 37504-364, Itajubá - MG, Brazil}

\author[0000-0002-4733-4994]{Joshua D. Simon}
\affiliation{The Observatories of the Carnegie Institution for Science, 813 Santa Barbara Street, Pasadena, CA 91101, USA}

\author{Andrew McWilliam}
\affiliation{The Observatories of the Carnegie Institution for Science, 813 Santa Barbara Street, Pasadena, CA 91101, USA}

\author[0000-0001-5761-6779]{Kevin C. Schlaufman}
\affiliation{William H.\ Miller III Department of Physics and Astronomy, Johns Hopkins
University, 3400 N Charles St, Baltimore, MD 21218, USA}

\author[0000-0002-5077-5774]{Paula Miquelarena}
\affiliation{Instituto de Ciencias Astron\'omicas, de la Tierra y del Espacio (ICATE), Espa\~na Sur 1512, CC 49, 5400 San Juan, Argentina}
\affiliation{Facultad de Ciencias Exactas, F\'isicas y Naturales, Universidad Nacional de San Juan, San Juan, Argentina}
\affiliation{Consejo Nacional de Investigaciones Cient\'ificas y T\'ecnicas (CONICET), Argentina}

\author[0000-0003-4749-8486]{Matias Flores Trivigno}
\affiliation{Instituto de Ciencias Astron\'omicas, de la Tierra y del Espacio (ICATE), Espa\~na Sur 1512, CC 49, 5400 San Juan, Argentina}
\affiliation{Facultad de Ciencias Exactas, F\'isicas y Naturales, Universidad Nacional de San Juan, San Juan, Argentina}
\affiliation{Consejo Nacional de Investigaciones Cient\'ificas y T\'ecnicas (CONICET), Argentina}

\author[0000-0002-8086-5746]{Marcelo Jaque Arancibia}
\affiliation{Instituto de Investigación Multidisciplinar en Ciencia y Tecnología, Universidad de La Serena, Raúl Bitrán 1305, La Serena, Chile}
\affiliation{Departamento de F\'isica y Astronom\'ia, Universidad de La Serena, Av. Cisternas 1200, La Serena, Chile}

\begin{abstract} 
Over the last decade, studies of large samples of binary systems identified chemical anomalies, and showed that they might be attributed to planet formation or planet engulfment. However, both scenarios have primarily been tested in pairs without known exoplanets. In this work, we explore these scenarios in the newly detected planet-hosting wide binary TOI-1173 A/B (projected separation $\sim 11,400$ AU) using high-resolution MAROON-X and ARCES spectra. We determined photospheric stellar parameters both by fitting stellar models and via the spectroscopic equilibrium approach. Both analyses agree and suggest that they are cool main sequence stars located in the thin disc. A line-by-line differential analysis between the components (B$-$A) displays an abundance pattern in the condensation temperature plane where the planet-hosting star TOI-1173 A is enhanced in refractory elements such as iron by more than 0.05 dex. This suggests the engulfment of $\sim$18 M$_{\oplus}$ of rocky material in star A. Our hypothesis is supported by the dynamics of the system \citep[detailed in our companion paper][]{Yana_Galarza:2024AJ....168...91G}, which suggest that the Super-Neptune TOI-1173 A $b$ might have been delivered to its current short period ($\sim7$ days) through circulatization and von Zeipel-Lidov-Kozai mechanisms, thereby triggering the engulfment of inner rocky exoplanets.

\end{abstract}

\keywords{Binary stars (154) --- Spectroscopy (1558) --- Stellar abundances (1577) --- Stellar atmospheres (1584) --- Fundamental parameters of stars (555) --- Solar analogs (1941) --- Exoplanets (498) --- Hot Neptunes (754)} 

\section{Introduction}\label{sec:intro}
It is assumed that the components of a binary system are both co-natal and co-eval, meaning they formed from the same parent material at approximately the same time. Consequently, due to this shared origin, they are expected to have a similar chemical composition. Studies of large samples of binary systems using high resolution ($R = \lambda / \Delta \lambda \geq 50,000$) and high signal-to-noise ratio (SNR $\geq$ 150 pixel$^{-1}$) spectra have reported homogeneity between components better than 0.05 dex in $\Delta$\feh\ \citep[e.g.,][]{Nelson:2021ApJ...921..118N}. However, there are also binary systems that do not follow either the co-natal or co-eval nature, exhibiting chemical inhomogeneities larger than 0.05 dex \citep[e.g.,][]{Hawkins:2020MNRAS.492.1164H, Nelson:2021ApJ...921..118N, Spina:2021NatAs...5.1163S, Yong:2023MNRAS.526.2181Y, Liu:2024Natur.627..501L}. Such inhomogeneities are also found among FGK-type stars when they are compared with the Sun, in particular in solar twins\footnote{Solar twins are stars with stellar parameters within \teff\ = 5777 $\pm$ 100 K, \logg\ = 4.44 $\pm$ 0.1 dex, and \feh\ = 0.0 $\pm$ 0.1 dex \citep{Ramirez:2014A&A...572A..48R}.}. In the last decade, several hypotheses have emerged to understand the origin of the chemical peculiarities observed in the Sun. One of them was proposed by \citet{Gustafsson:2018A&A...616A..91G, Gustafsson:2018A&A...620A..53G}, suggesting that the chemical makeup of the solar photosphere might have been affected by radiative dust cleansing in the primordial nebula. \citet{Adibekyan:2014A&A...564L..15A} proposed that Galactic chemical evolution (GCE) produces trends in the abundances of elements as a function of time, resulting in chemical anomalies when stars of different ages are compared with the Sun. \citet{Ramirez:2019MNRAS.490.2448R} suggested that the chemical abundances of binary systems might be attributed to inhomogeneities in the molecular clouds from which each component forms. This hypothesis also applies to solar twins and the Sun, as they are not formed from the same molecular cloud. Other scenarios include planetary locking and planetary engulfment. Both are the focus of this work, as they have been extensively tested in binary systems \citep[e.g.,][]{Ramirez:2011ApJ...740...76R, Tucci:2014ApJ...790L..25T, Teske:2016ApJ...819...19T, Saffe:2017A&A...604L...4S,  Fan_Liu:2021MNRAS.508.1227L, Yana_Galarza:2021ApJ...922..129G, Spina:2021NatAs...5.1163S, Flores:2024MNRAS.52710016F, Liu:2024Natur.627..501L, Miquelarena:2024arXiv240606705M}. These scenarios are discussed in detail throughout the paper. 

In the context of rocky planet locking, exoplanets can modify star surfaces by sequestering refractory elements during planet formation. As a result, stars that host rocky planets are expected to be deficient in refractory elements. \citet{Melendez:2009ApJ...704L..66M} proposed this scenario when they found that the Sun's refractory elements are deficient compared to the average composition of 11 solar twin stars in the condensation temperature plane\footnote{In this work, the condensation temperature plane refers to the comparison of the chemical abundances of a star as a function of their condensation temperature.}. However, more recent studies, based on disc simulations, suggest that giant exoplanets might also be responsible for the depletion of refractories in the Sun \citep{Booth:2020MNRAS.493.5079B, Huhn:2023A&A...676A..87H}. In these simulations, the formation of a giant gas planet opens a gap in the protoplanetary disc, trapping the refractory elements and preventing them from falling into the star.

In the planet engulfment scenario, the ingestion of a rocky exoplanet could substantially enhance the refractory elements on the surface of stars. One might also observe an enhancement in the lithium surface abundance, depending on the timing of the engulfment \citep[e.g.,][]{Melendez:2017A&A...597A..34M, Yana_Galarza:2021ApJ...922..129G, Flores:2024MNRAS.52710016F}. 

Over the last two decades, several studies have tested both planet formation and planet engulfment scenarios, mainly using solar-type stars with and without exoplanets. However, as of today, convincing evidence to confirm either of those scenarios is still lacking \citep[e.g.,][]{Ramirez:2009A&A...508L..17R, Gonzales_Hernandez:2010ApJ...720.1592G, Schuler:2011ApJ...732...55S, Bedell:2018ApJ...865...68B, Yana:2021MNRAS.502L.104Y}.

Twin planet-hosting binary systems, namely components of a pair with similar stellar properties, are likely the key to understanding whether planets can alter the surface of stars. This relies on the assumption that pairs are co-eval and co-natal, meaning that any anomalies detected in their components might be associated with the presence of one or more exoplanets. As of today, the chemical abundances of only 37 planet-hosting wide binaries (WBs) have been determined in order to detect planet signatures, namely planet formation or planet engulfment \citep{Fan_Liu:2014MNRAS.442L..51L, Mack:2014ApJ...787...98M, Teske:2015ApJ...801L..10T, Ramirez:2015ApJ...808...13R, Biazzo:2015A&A...583A.135B, Teske:2016ApJ...819...19T, Saffe:2017A&A...604L...4S, Saffe:2015A&A...582A..17S, Fan_Liu:2018A&A...614A.138L, Ramirez:2011ApJ...740...76R, Tucci:2019A&A...628A.126M, Fan_Liu:2021MNRAS.508.1227L, Saffe:2019A&A...625A..39S, Emi_Jofre:2021AJ....162..291J, Behmard:2023MNRAS.521.2969B, Flores:2024MNRAS.52710016F}. Unfortunately, the results remain inconclusive due to conflicting findings from different studies. In particular, the planet engulfment scenario is considered less favorable, as pointed out by \citet{Behmard:2023MNRAS.521.2969B}, whose findings reveal that the origin of the chemical anomalies is likely primordial. This contradicts the planet engulfment scenario proposed by \citet{Spina:2018MNRAS.474.2580S}, \cite{Yong:2023MNRAS.526.2181Y} and \citet{Liu:2024Natur.627..501L}, aimed at explaining the inhomogeneities observed in WBs, albeit not all stars in their sample harbor exoplanets.

The detection of planet signatures is a challenging task that relies on the precise determination of stellar parameters such as effective temperature (\teff), metallicity (\feh), surface gravity (\logg), age, mass, rotation and magnetic activity. The obtainable precision strongly depends on the quality of the spectra and the selection of a good line list, with well measured (laboratory measurements) atomic data (e.g., excitation potential and $\log{gf}$). It has been demonstrated that the analysis of high resolution ($\geq$ 60,000) and high SNR ($\geq$ 200) spectra results in better precision not only in stellar parameters, but also when deriving stellar chemical abundances ($< 0.05$ dex). Additionally, the analysis of dynamical timescales should be considered in planet-hosting wide binaries to potentially detect planet engulfment triggered by planet migration.

Precision in a chemical analysis significantly increases when employing the differential technique, with a precision as good as $\sim$0.01-0.02 dex achievable for solar twins \citep[e.g.,][]{Melendez:2009ApJ...704L..66M, Nissen:2015A&A...579A..52N, Spina:2016A&A...593A.125S, Bedell:2018ApJ...865...68B, Yana_Galarza:2021ApJ...922..129G}, thus opening up the possibility of detecting planet-related signatures on those objects. A caveat is that the achievable precision decreases if there are large differences between the stars. In particular, differences in metallicity and effective temperatures are the largest contributors to a precision decrease \citep[e.g.,][]{Henrique:2017CaJPh..95..855R}. From a physical perspective, the decrease in precision mostly come from large enough differences in continuum opacities (which in a differential abundance analysis are assumed to be zero). For differences in effective temperatures larger than about $500$ K and metallicity differences on the order of $0.15$ dex, non-LTE effects are also an important source of differences (which is not our case). 
From a data analysis perspective, effective temperature differences also contribute to differences during normalization, and continuum placement during line measurements can play an important role in the differential analysis. Any effect that changes the line profile, such as rotation, can influence line-measurements and decrease achievable precision. It is also important to highlight that differences in surface gravity must be larger than the differences observed in our pair to have any effect on the achievable precision, as the changes caused in line profiles are mostly observable in saturated and resonant lines, which we avoid in our analysis. Despite these caveats, as we show in Section \ref{sec:stellaparam}, the differences in the stellar parameters of TOI-1173 A/B are sufficiently small that a differential abundances brings important improvements to the elemental abundance precision. Additional extensive discussions on the benefits of using the differential approach in FGK-type stars are provided in \citet{Smiljanic:2007A&A...468..679S, Ramirez:2011ApJ...743..135R, Feltzing:2013NewAR..57...80F, Nissen:2018A&ARv..26....6N} and \citet{Jofre:2019ARA&A..57..571J}.

Therefore, in this paper, we adhere to the differential analysis approach to investigate whether the origin of the chemical anomalies detected in the planet-hosting wide binary TOI-1173 A/B can be attributed to planet signatures. In our companion paper \citep{Yana_Galarza:2024AJ....168...91G}, we report the discovery of a Super-Neptune in the component TOI-1173 A, thereby expanding the restricted sample of planet-hosting WBs. Our pair represents the second WB system discovered to host planets with separations greater than 10,000 AU, following the system HAT-P-4 A with a separation of $\sim$29,500 AU \citep{Mugrauer:2014MNRAS.439.1063M, Saffe:2017A&A...604L...4S}.

In Section \ref{sec:obs}, we describe the target selection, observations, and data reductions. Section \ref{sec:stellaparam} presents the determination of stellar parameters, age, and masses, while Section \ref{sec:chem} discusses the inferred chemical composition. In Section \ref{sec:discussions}, we present the discussion of our results. Section \ref{sec:conclusions} provides a summary and conclusions. Finally, Appendix \ref{appendix} describes the GCE corrections applied to the binary system, when compared to the Sun.

\begin{table}
    \begin{scriptsize}
    \centering
    \begin{threeparttable}
	\caption{Photometric parameters and Galactic orbital and space velocities of TOI-1173 A/B. The errors in \textit{Gaia} magnitudes were adopted from Vizier \citep{Ochsenbein:2000A&AS..143...23O}.} 
    \begin{tabular}{lcc}
    \hline 
	\hline
    Parameter & TOI-1173A  & TOI-1173B   \\
    \hline
    \textit{Gaia} DR3 $G$$^{a}$       & $10.804\pm0.003$  & $11.389\pm0.003$ \\
    \textit{Gaia} DR3 $(B-P)$$^{a}$   & $11.208\pm0.003$  & $11.856\pm0.003$ \\
    \textit{Gaia} DR3 $(R-P)$$^{a}$   & $10.236\pm0.004$  & $10.761\pm0.004$ \\
    \textit{Gaia} DR3 parallax$^{a}$  & $7.566\pm0.012$ & $7.565\pm0.016$ \\
    RUWE$^{a}$                        & 0.839             & 0.881             \\
    Radial velocity$^{a}$ (km s$^{-1}$) & $-45.920\pm0.349$ & $-45.992\pm0.553$  \\
    $\mu_{\alpha}^{\ast a}$ (mas yr$^{-1}$)  & $57.786\pm0.011$  & $57.419\pm0.015$ \\
    $\mu_{\delta}^{a}$ (mas yr$^{-1}$)                  & $-92.392\pm0.012$ & $-93.021\pm0.017$ \\
    2MASS $J$$^{b}$                    & $9.564\pm0.020$   & $10.043\pm0.022$ \\ 
    2MASS $H$$^{b}$                    & $9.228\pm0.017$   & $9.598\pm0.023$ \\ 
    2MASS $K_{\text{s}}$$^{b}$         & $9.134\pm0.015$   & $9.495\pm0.020$ \\ 
    E(B-V)$^{\gamma}$                      & \multicolumn{2}{c}{$0.010\pm0.014$}  \\
    Distance$^{\delta}$ (pc)               & $131.810^{+0.193}_{-0.199}$ & $131.544^{+0.342}_{-0.252}$ \\
    Projected separation                   & \multicolumn{2}{c}{86.735 arcsec,  11432.592 AU}  \\
    $z_{\rm{max}}$          (kpc)          & 0.192              & 0.195 \\
    Eccentricity                           & 0.277              & 0.277 \\
    Perigalacticon           (kpc)         & 5.725              & 5.710 \\
    Apogalacticon           (kpc)          & 10.104             & 10.092 \\
    $U$           (km s$^{-1}$)            & 72.282             & 72.284 \\
    $V$           (km s$^{-1}$)            & 39.120             & 39.559 \\
    $W$           (km s$^{-1}$)            & 3.882              & 4.153 \\
    3D velocity difference (km s$^{-1}$)   & \multicolumn{2}{c}{$0.516$}  \\
    \hline
    \end{tabular}
    \begin{tablenotes}
    \item $^{a}$\cite{GaiaDR3:2023A&A...674A...1G}, $^{b}$\cite{2mass:2006AJ....131.1163S}, $^{\gamma}$\cite{Lallement:2014A&A...561A..91L, Lallement:2018A&A...616A.132L}, $^{\delta}$\cite{Bailer:2021AJ....161..147B}.
    \end{tablenotes}
    \label{tab:parameters}
    \end{threeparttable}
    \end{scriptsize}
\end{table}

\section{Observations} 
\subsection{Target Selection}
\label{sec:obs}
The wide binary system TOI-1173 A/B (\textit{Gaia} DR3 1686171213716517504, and \textit{Gaia} DR3 1686170870119133440 respectively), was selected using the color constraints of \citet[][see their Table 1]{Yana_Galarza:2021MNRAS.504.1873Y} originally designed to identify FGK-type stars in the $Gaia$ database \citep{GaiaDR3:2023A&A...674A...1G} but now applied to the one million binaries catalog of \citet{El-Badry:2021MNRAS.tmp..394E}. The components of the pair share identical radial velocities (RVs), and their \textit{Gaia} Renormalized Unit Weight Error (RUWE, \citealp{GaiaDR3:2023A&A...674A...1G}), which is an indicator of multiplicity, are lower than 1 (see Table \ref{tab:parameters}), indicating that the components are not themselves binary systems \citep{Kervella:2022A&A...657A...7K}. We corrected the systematic offsets in the \textit{Gaia} DR3 parallaxes following \citet{Lindegren:2021A&A...649A...4L}. The distance was computed by simply inverting the parallax, resulting in $\sim132$ and pc for both components, in agreement with the distances inferred by \cite{Bailer:2021AJ....161..147B}. The projected separation of the pair was initially estimated as 11,463.25 AU by \citet{El-Badry:2021MNRAS.tmp..394E}. Using the distance from \cite{Bailer:2021AJ....161..147B}, we obtained a similar separation ($\sim$11,400 AU).

\begin{figure}
 \includegraphics[width=\columnwidth]{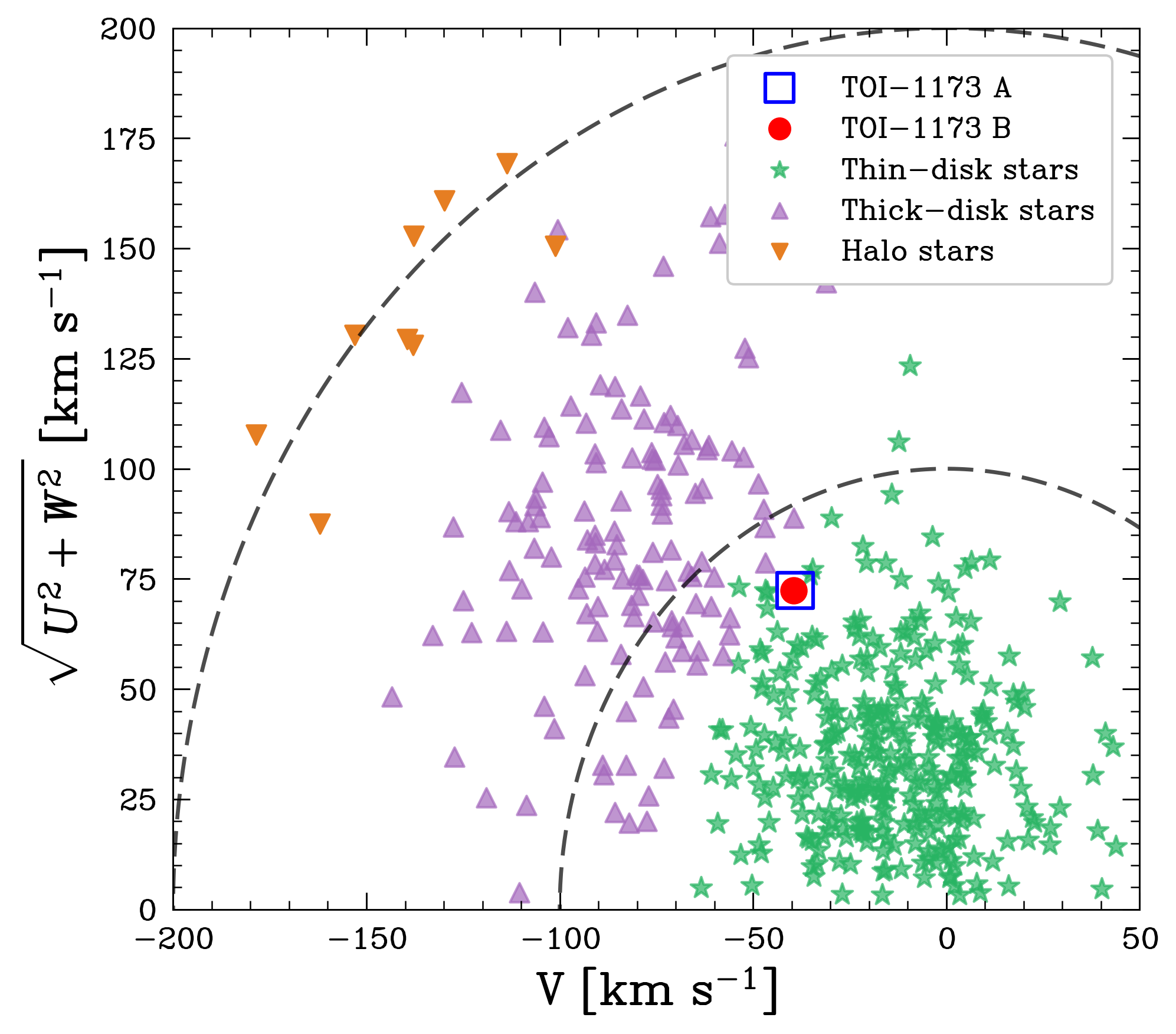}
 \centering
 \caption{Toomre diagram for the binary system TOI-1173 A/B indicating that our pair is located in the thin-disk. The stars, triangles and inverted triangles symbols represent stars from the thin-disk, thick-disk, and halo of the Galaxy, respectively, for context. Data taken from \citet{Ramirez:2007A&A...465..271R}.}
 \label{fig:Toomre_diagram}
\end{figure}

We determined Galactic orbits using the \textsc{Gala}\footnote{\url{https://github.com/adrn/gala}} code \citep{gala, adrian_price_whelan_2020_4159870} and the \textit{Gaia} DR3 astrometry \citep{GaiaDR3:2023A&A...674A...1G}. In brief, we first transformed the astrometric data and radial velocities of the pair into galactocentric Cartesian units using \textsc{Astropy} \citep{astropy:2022}, assuming that the Sun's position and velocity are $x_{\odot} = (-8.3, 0, 0)$ kpc and $v_{\odot} = (-11.1, 244, 7.25)$ km s$^{-1}$ \citep{Schonrich:2010MNRAS.403.1829S, Schonrich:2012MNRAS.427..274S}. Then, we used {\sc Gala} to carry out the orbital integration with the default potential {\sc MilkyWayPotential}. This potential is a mass model for the Milky Way consisting of a spherical nucleus and bulge \citep{Hernquist:1990ApJ...356..359H}, a Miyamoto–Nagai disc \citep{Miyamoto:1975PASJ...27..533M, Bovy:2015ApJS..216...29B}, and a spherical Navarro-Frenk-White (NFW) dark matter halo \citep{Navarro:1996ApJ...462..563N}. We used a timestep of 1 Myr and integrate for 4 Gyr, corresponding to $\sim$20 orbits for each component of the pair.

Our results suggest that both components have similar maximum vertical excursion ($z_{\rm{max}}$), eccentricity, perigalacticon and apogalacticon values. The Galactic space velocities ($U, V, W$) place the wide binary within the thin-disc kinematic distribution on the Toomre diagram ($V$ versus $\sqrt{U^{2}+V^{2}}$; see Fig. \ref{fig:Toomre_diagram}). The Galactic orbits and space velocities are listed in Table \ref{tab:parameters}. We estimated the 3D velocity difference ($\Delta v_{\rm{3D}}$) to be 0.52 km s$^{-1}$. According to simulations by \citet{Kamdar:2019ApJ...884..173K}, components with $\Delta v_{\rm{3D}} < 2$ km s$^{-1}$ and separations below $\sim 10^{7}$ AU have a high probability of being conatal. Therefore, given the $\sim 11,400$ AU separation and a $\Delta v_{\rm{3D}}$ below 1 km s$^{-1}$, we conclude that TOI-1173 A/B is truly conatal.

The component A of the binary system hosts a Super-Neptune exoplanet. TESS observed TOI-1173 A in 9 sectors (14, 15, 21, 22, 41, 47, 48, 74, and 7). As we will describe in subsection \ref{sub:spectroscopic_observations}, we also obtained MAROON-X spectra for the A component, and using both the transit and radial velocity observations, we infer precise exoplanet parameters, which are listed in Table \ref{tab:exoplanet_parameters}. The mass and radius of TOI-1173 A $b$ yield a density of $\sim0.195 \pm 0.018$ g cm$^{-3}$, making it the first puffy Super-Neptune discovered in a wide binary system \citep{Yana_Galarza:2024AJ....168...91G}. Inflated exoplanets represent a challenge for current models of exoplanet formation. As for hot Jupiters, it is believed that puffy exoplanets might not have formed in situ. Instead, there is a consensus that they might have undergone migration. In 
\cite{Yana_Galarza:2024AJ....168...91G} we discussed the planetary dynamical history in detail, and the migration of this puffy planet constitutes a crucial component of our planet engulfment hypothesis.

\begin{table}
    \centering
	\caption{Exoplanet parameters of TOI-1173 A $b$. Data taken from \citet{Yana_Galarza:2024AJ....168...91G}.} 
    \begin{tabular}{lcr}
    \hline 
	\hline
    Parameter          & TOI-1173 A $b$             &  Units \\
    \hline
    Orbital period ($P_b$)   & $7.06466_{-0.00029}^{+0.00028}$  &  days \\
    Semi-amplitude ($K_b$) & $9.67_{-0.46}^{+0.47}$ & m s$^{-1}$ \\
    Semi-major axis ($a_\mathrm{b}$) & $0.0696\pm0.0014$ &  AU \\
    Eccentricity ($e_b$) & $0.023_{-0.016}^{+0.025}$ & \\
    Mass ($M_\mathrm{b}$) & $27.4\pm1.7$ & $\mathrm{M_{\oplus}}$ \\
    Radius ($R_\mathrm{b}$) & $9.19\pm0.18$ & $\mathrm{R_{\oplus}}$ \\
    Density ($\rho_\mathrm{b}$) & $0.195_{-0.017}^{+0.018}$ & g cm$^{-3}$ \\
    \hline
    \hline
    \end{tabular}
    \label{tab:exoplanet_parameters}
\end{table}

\subsection{Spectroscopic Observations}\label{sub:spectroscopic_observations}
TOI-1173 A and B were observed under the program ID GN-2022A-Q-227 with the high-resolution echelle spectrograph MAROON-X \citep{Seifahrt:2018SPIE10702E..6DS, Seifahrt:2022SPIE12184E..1GS} mounted on the 8.1 m Gemini North Telescope of The International Gemini Observatory located in Hawaii. The instrument has a spectral resolution $R\sim$85,000 and possesses two optical-NIR channels, with wavelength coverage of 500-670 nm and 650-900 nm for the blue and red channels, respectively. We obtained ten spectra for TOI-1173 A, three for TOI-1173 B, and one for the Sun. The latter was acquired through the reflected light of the Vesta asteroid. The data reduction was performed by the MAROON-X instrument team using a custom {\sc python3} pipeline that provides optimally extracted and wavelength-calibrated 1D spectra. The radial velocities were calculated using the SpEctrum Radial Velocity AnaLyser (SERVAL, \citealp{Zechmeister:2018A&A...609A..12Z}) pipeline, which creates high signal-to-noise templates by coadding the observed spectra. The templates are then shifted for radial velocity and compared to each individual observation employing least-squares fitting.

We also obtained spectra for the pair using ARCES on the 3.5 m telescope at Apache Point Observatory. A solar spectrum was taken by observing the sky during twilight. Calibration frames, such as biases and flats, were taken in the afternoon of the observations. All observations were conducted in May 2023. The ARCES spectrograph provides spectra with a resolution of $R=31,500$, covers the entire visible wavelength range from 320 nm to 1000 nm. Reduction of the ARCES data was done via the \textsc{CERES}\footnote{\url{https://github.com/rabrahm/ceres}} package. It conducts bias subtraction (via a master bias calculated as the median value of a series of bias frames observed in the afternoon). Orders are traced with a master flat (calculated as the median value for a series of flats taken in the afternoon calibrations). The master flat, which combines the blue and red channels, is also used for flat-field correction of the science frames. Wavelength calibration relies on a ThAr lamp exposure taken immediately after the science frames. For more details on ARCES data reduction with CERES, we refer the reader to \cite{CERES:2017PASP..129c4002B}. 

\begin{table*}
    \centering
	\caption {Stellar parameters of TOI-1173 A/B. $^{(\star)}$ Photometric effective temperature estimated with \textsc{colte} \citep{Casagrande:2021MNRAS.507.2684C}. $^{(\dagger)}$Trigonometric surface gravity calculated following \citet{Yana_Galarza:2021MNRAS.504.1873Y}. $^{(\ddagger)}$ Stellar parameters of B determined using \qq\ \citep{Ramirez:2014A&A...572A..48R}. $^{(\times)}$ Stellar parameters of B determined using \textsc{xiru} \citep{Alencastro_thesis:1885-288051}. (B$-$A) refers to the stellar parameters of the B component relative to the A component.}
	\begin{tabular}{lcccccc} 
		\hline
		\hline
		ID             & \teff           & \logg             & \feh                & \vmic             & \teff $^{(\star)}$ & \logg $^{(\dagger)}$ \\     
		               & (K)             & (dex)             & (dex)               &  (km s$^{-1}$)    &   (K)              &  (dex)               \\             
        \hline
        \multicolumn{7}{c}{\textsc{Qoyllur-Quipu} (\qq) - The Sun as reference} \\
        \hline                                                                                                                                                       
		TOI-1173 A (A$-$Sun) &  5300 $\pm$ 9   & 4.330 $\pm$ 0.022 &  0.117  $\pm$ 0.010 & 0.67  $\pm$ 0.03  & 5367 $\pm$ 50    & 4.423 $\pm$ 0.022    \\ 
		TOI-1173 B (B$-$Sun) &  4966 $\pm$ 20  & 4.280 $\pm$ 0.022 &  0.074  $\pm$ 0.016 & 0.40  $\pm$ 0.11  & 5068 $\pm$ 43    & 4.477  $\pm$ 0.023   \\  
		\hline
		 \multicolumn{7}{c}{\textsc{xiru} - The Sun as reference} \\
        		\hline                                                                                                                                                      
		TOI-1173 A (A$-$Sun) &  5315 $\pm$ 9   & 4.439 $\pm$ 0.052 &  0.126  $\pm$ 0.010 & 0.77  $\pm$ 0.02  & 5369 $\pm$ 51    &  4.419 $\pm$ 0.022   \\ 
		TOI-1173 B (B$-$Sun) &  4995 $\pm$ 15  & 4.451 $\pm$ 0.107 &  0.107  $\pm$ 0.020 & 0.59  $\pm$ 0.03  & 5075 $\pm$ 43    &  4.480 $\pm$ 0.022   \\  
		\hline
		 \multicolumn{7}{c}{\textsc{isochrones}} \\     
		 \hline 
		TOI-1173 A  (A$-$Sun)    &  5350 $\pm$ 34  & 4.450 $\pm$ 0.020 &  0.139  $\pm$ 0.065 & 1.11  $\pm$ 0.01  & 5370 $\pm$ 50    &  4.444 $\pm$ 0.024   \\ 
		TOI-1173 B  (B$-$Sun)    &  5047 $\pm$ 34  & 4.530 $\pm$ 0.030 &  0.114  $\pm$ 0.069 & 1.09  $\pm$ 0.01  & 5078 $\pm$ 43    &  4.520 $\pm$ 0.025   \\
		\hline
		 \multicolumn{7}{c}{Differential approach between components using A as reference} \\     
		 \hline 
		TOI-1173 B (B$-$A)$^{\ddagger}$      &  5054 $\pm$ 8  & 4.490 $\pm$ 0.020 &  $-$0.062  $\pm$ 0.007 & 0.99  $\pm$ 0.03  & \multicolumn{2}{c}{Reference star A from {\sc isochrones}}  \\	
		TOI-1173 B (B$-$A)$^{\times}$        &  4983 $\pm$ 8   & 4.390 $\pm$ 0.022 &  $-$0.050  $\pm$ 0.006 & 0.55  $\pm$ 0.04 & \multicolumn{2}{c}{Reference star A from {\sc xiru}}  \\  
		\hline
		\hline
	\end{tabular}
	\label{tab:fundamental_parameters}
\end{table*}

\subsection{Data Treatment}
We used the \textsc{iSpec}\footnote{\url{https://www.blancocuaresma.com/s/iSpec}} \citep{Blanco:2014A&A...569A.111B, Blanco:2019MNRAS.486.2075B} tool to carry out radial/barycentric velocity correction only for the ARCES spectra. For the MAROON-X spectra, we employed the radial velocity provided by the MAROON-X instrument team. Once the spectra were shifted to the rest frame, we normalized each individual spectrum with the {\sc continuum} task in \textsc{IRAF}\footnote{\textsc{IRAF} is distributed by the National Optical Astronomical Observatories, which is operated by the Association of Universities for Research in Astronomy, Inc., under a cooperative agreement with the National Science Foundation.} by fitting the spectra preferably with third degree spline functions in the blue orders, and low-order polynomials in the red orders. The co-addition of the spectra is done after normalization, using the IRAF {\sc scombine} task. It resamples each spectrum and combines them using the average values, therefore increasing achievable SNR as well. The resulting MAROON-X spectra have signal-to-noise ratios of $\sim$540, $\sim$300, and $\sim$200 pixel$^{-1}$ at $\sim$630 nm for the Sun, TOI-1173 A, and TOI-1173 B, respectively. For the combined ARCES spectra, the SNR is 140 pixel$^{-1}$ for the A component and 150 pixel$^{-1}$ for the B component, both at $\sim$665 nm. 

\section{Fundamental Parameters} \label{sec:stellaparam}
\subsection{Equivalent widths and stellar parameters}
We measured the equivalent widths (EWs) of stellar absorption lines by fitting Gaussians to the line profiles using the \textsc{KAPTEYN}\footnote{\url{https://www.astro.rug.nl/software/kapteyn/index.html}} kmpfit package \citep{KapteynPackage}. The line-by-line measurements were carried out carefully choosing local pseudo-continuum regions of 6 \AA, with the line of interest located at the center. This procedure was performed separately for components A and B, with the Sun as the reference star, and then for component B with A as the reference. The line list used in this work is an updated version of the line list first presented by \citet{Melendez:2014ApJ...791...14M} and includes information about excitation potentials, oscillator strengths, and laboratory $\log{gf}$ values. We include hyperfine structure and isotopic contributions from \cite{McWilliam:1998AJ....115.1640M,Prochaska2000a,Prochaska2000b,klose2002,Cohen:2003ApJ...588.1082C,Blackwell-Whitehead2005a,Blackwell-Whitehead2005b,Lawler2014}, and from the Kurucz\footnote{\url{http://kurucz.harvard.edu/linelists.html}} line lists. To avoid saturation effects, only iron lines with measured $EW < 130$ m\AA\ were used in the determination of stellar parameters.

The spectroscopic stellar parameters, namely the effective temperature (\teff), surface gravity (\logg), metallicity (\feh), and microturbulent velocity (\vmic) are determined via spectroscopic equilibrium, satisfying three conditions: (i) The excitation potential equilibrium to get the \teff; (ii) Ionization equilibrium  (same iron abundance from 2 ionization stages) to get the \logg, and (iii) Non-dependence of the differential abundance and the reduced equivalent width (log($EW/\lambda$)) to get the \vmic. The spectroscopic equilibrium is based on measurements of iron abundances (\ion{Fe}{1} and \ion{Fe}{2}). In a line-by-line differential analysis, the iron abundance is measured relative to a reference star, usually the Sun. This technique minimizes the impact of model atmospheres as well as errors in laboratory atomic data (e.g., oscillator strengths) since they cancel out in each line calculation. Precision is maximized when both the reference star and the target star are twins. Using solar twins, this method typically achieves a precision of about 10 K in \teff, 0.01 dex in \logg\ and 0.01 dex in \feh\ \citep[e.g.,][]{Ramirez:2014A&A...572A..48R, Bedell:2018ApJ...865...68B, Yana_Galarza:2021MNRAS.504.1873Y}.

In line with our previous works, we utilized the automatic Python code Qoyllur-quipu\footnote{\url{https://github.com/astroChasqui/q2}} (\qq) to determine stellar parameters through spectroscopic equilibrium. A detailed explanation of \qq's functionality and performance can be found in \cite{Ramirez:2014A&A...572A..48R}. We configured \qq\ to employ the Kurucz \textsc{ODFNEW} model atmospheres \citep{Castelli:2003IAUS..210P.A20C} and the 2019 local thermodynamic equilibrium (LTE) code \textsc{moog} \citep{Sneden:1973PhDT.......180S}. The stellar parameters of the pair relative to the Sun and to each other are reported in Table \ref{tab:fundamental_parameters}. As a sanity check, we estimated photometric \teff\ using the Colour-\teff\ routine \textsc{colte}\footnote{\url{https://github.com/casaluca/colte}}, which uses \textit{Gaia} and 2MASS photometry in the InfraRed Flux Method (IRFM) to establish color-effective temperature relations sensitive to \feh\ and \logg. The IRFM method and the routine are well described in \cite{Casagrande:2021MNRAS.507.2684C}. Additionally, we also computed \logg\ through the trigonometric parallax method, as it is a crucial input parameter in the isochrone fitting method for inferring ages. The trigonometric \logg\ is calculated using Equation (3) from \cite{Yana_Galarza:2021MNRAS.504.1873Y}, with \textit{Gaia} DR3 parallaxes, Johnson $V$ magnitudes \citep{Kharchenko:2001KFNT...17..409K} corrected for reddening using E(B-V) = 0.01 \citep{Lallement:2018A&A...616A.132L}, bolometric corrections from \cite{Melendez:2006ApJ...641L.133M}, and stellar masses from isochrone fitting (see subsection \ref{sec:ages}). 

Table \ref{tab:fundamental_parameters} shows an agreement between the photometric and spectroscopic \teff\ for the A component, but not for the B component, where the difference is $\sim$100 K. There is a notable difference between the trigonometric and spectroscopic \logg\ for both stars, with a deviation of 0.093 dex for star A and 0.197 dex for star B. Such disparities in estimating stellar parameters are not surprising, as the spectroscopic method, especially ionization equilibrium, relies on different physical assumptions, the quality of the spectra (high SNR and $R$), and the number of reliable \ion{Fe}{1} and \ion{Fe}{2} lines with accurate EWs measurements. 

In our analysis, although the MAROON-X spectra are of high quality, there are only between 10 to 11 clean \ion{Fe}{2} lines for each component. This limitation could potentially affect the ionization equilibrium, resulting in incorrect \logg\ values. On the other hand, as mentioned earlier, the differential analysis technique is more accurate when the reference star is similar to the target star \citep[e.g.,][]{Henrique:2017CaJPh..95..855R}. This is not the case for our binary system, so the continuum and model atmosphere could influence the excitation/ionization balance of iron lines. It is known that in cool stars, particularly K-type stars, finding the continuum below 500 nm is challenging. To minimize the impact of continuum normalization on the B component, which is the cooler star, we normalized only orders with wavelengths greater than 500 nm. Additionally, we determined the stellar parameters using absolute iron abundances; however, we found similar deviations in \logg.

In their study, \citet{Delgado_Mena:2017A&A...606A..94D} analyzed a sample of FGK stars from the HARPS GTO program and observed significant discrepancies between spectroscopic and trigonometric \logg\ values. They found an average difference of $-0.22$ ($\pm 0.10$) dex for cool stars (\teff\ $<$ 5200 K) and 0.21 ($\pm 0.13$) dex for hot stars (\teff\ $>$ 6100 K), as shown in their Fig. 2. To address this issue, the authors applied corrections based on linear fits to the difference in \logg\ versus \teff. The corrected results were more realistic and, as emphasized by the authors, aligned with isochrones in the Kiel diagram (\logg\ vs. \teff). Following \citet{Delgado_Mena:2017A&A...606A..94D}, and assuming that the trigonometric \logg\ represents the more realistic value for cool stars, we run \qq\ by fixing the trigonometric \logg\ value. Nonetheless, we found significant trends in the excitation balance, reduced EW, and differences between iron abundances inferred from \ion{Fe}{1} and \ion{Fe}{2}. Thus, the spectroscopic equilibrium cannot be satisfied in any way.

The origin of the discrepancies between trigonometric and spectroscopic surface gravities is still unknown. \citet{Bensby:2014A&A...562A..71B} studied the impact of NLTE effects on \ion{Fe}{1} for stars with \teff\ between 4600 and 6800~K and reported significant changes in the derived \logg\ for stars with \teff\ $>$ 6000~K when NLTE effects are included (see their Fig. 6). No significant dependence of the surface gravity on the NLTE correction was found for cooler stars (\teff $<$ 5500 K), indicating that NLTE is unlikely to be the origin of the disagreement between the spectroscopic and trigonometric \logg\ in TOI-1173 A/B. These discrepancies are not new. They have already been reported in other studies \citep[e.g.,][]{Mortier:2013A&A...558A.106M, Bensby:2014A&A...562A..71B, Delgado_Mena:2014A&A...562A..92D} as a trend of \logg$_{\rm{spec}}$ $-$ \logg$_{\rm{trig}}$ vs. effective temperature, even when \logg\ \ is estimated from photometric light curves  \citep[see Fig. 1 in][]{Tsantaki:2014A&A...570A..80T}.

Delving deeper in our analysis, we used the code \textsc{xiru}\footnote{\url{https://github.com/arthur-puls/xiru}} \citep{Alencastro_thesis:1885-288051} to investigate inconsistences between trigonometric and spectroscopic \logg. Similar to \qq, \textsc{xiru} is also a Python code written to determine stellar parameters through the spectroscopic equilibrium using \textsc{MOOG} and Kurucz \textsc{ODFNEW} model atmosphere. However, the methodology is different as \textsc{xiru} uses an implementation based on Broyden's method \citep{Broyden1965ACO}, widely used to solve systems of nonlinear equations by iteratively updating a Jacobian matrix to achieve convergence. As seen in Table \ref{tab:fundamental_parameters}, the stellar parameters obtained with \textsc{xiru} are similar to those from \qq, except for \logg, which more closely aligns with the trigonometric values. However, the spectroscopic \teff\ values are slightly underestimated compared to the photometric \teff. \textsc{xiru} provides better \logg\ values than \qq\ for cool stars, and the underestimation in \logg\ computed with \qq\ remains unclear. 

We also employed the \textsc{Isochrones} package\footnote{\url{https://github.com/timothydmorton/isochrones}} \citep{morton2015} to estimate the stellar parameters for the pair. In brief, the \textsc{isochrones} package fits MESA Isochrones and Stellar Tracks \cite[MIST;][]{dotter2016,choi2016,paxton2011,paxton2013,paxton2015,paxton2018,paxton2019} using \textsc{MultiNest}\footnote{\url{https://ccpforge.cse.rl.ac.uk/gf/project/multinest/}} \citep{feroz2008,feroz2009,feroz2019} via \textsc{PyMultinest} \citep{buchner2014}. We fitted \textit{Gaia} DR3 G band magnitude \citep{arenou2018,evans2018,hambly2018, gaia2021,fabricius2021,Lindegren:2021A&A...649A...4L,lindegren2021b,torra2021}, and $J$, $H$, and $K_{\rm{s}}$ bands from the Two Micron All Sky Survey (2MASS) All-Sky Point Source Catalog \citep[PSC,][]{skrutskie2006}. Additionally, and we include the \textit{Gaia} DR3 parallaxes of our targets. The extinction $A_V$ was inferred based on three-dimensional (3D) maps of extinction in the solar neighborhood from the Structuring by Inversion the Local Interstellar Medium (Stilism\footnote{\url{https://stilism.obspm.fr/}}) program \citep{lallement2014,lallement2018,capitanio2017}, and the spectroscopically-informed metallicity in our prior (within $\feh\ = \pm 0.1$ dex). For more details on our methodology, we refer the reader to \cite{Henrique:2022AJ....163..159R}. The resulting stellar parameters are in excellent agreement with the photometric and trigonometric values, with an average difference in \teff\ and \logg\ of only 26 K and 0.01 dex, respectively.

\begin{figure}
 \includegraphics[width=\columnwidth]{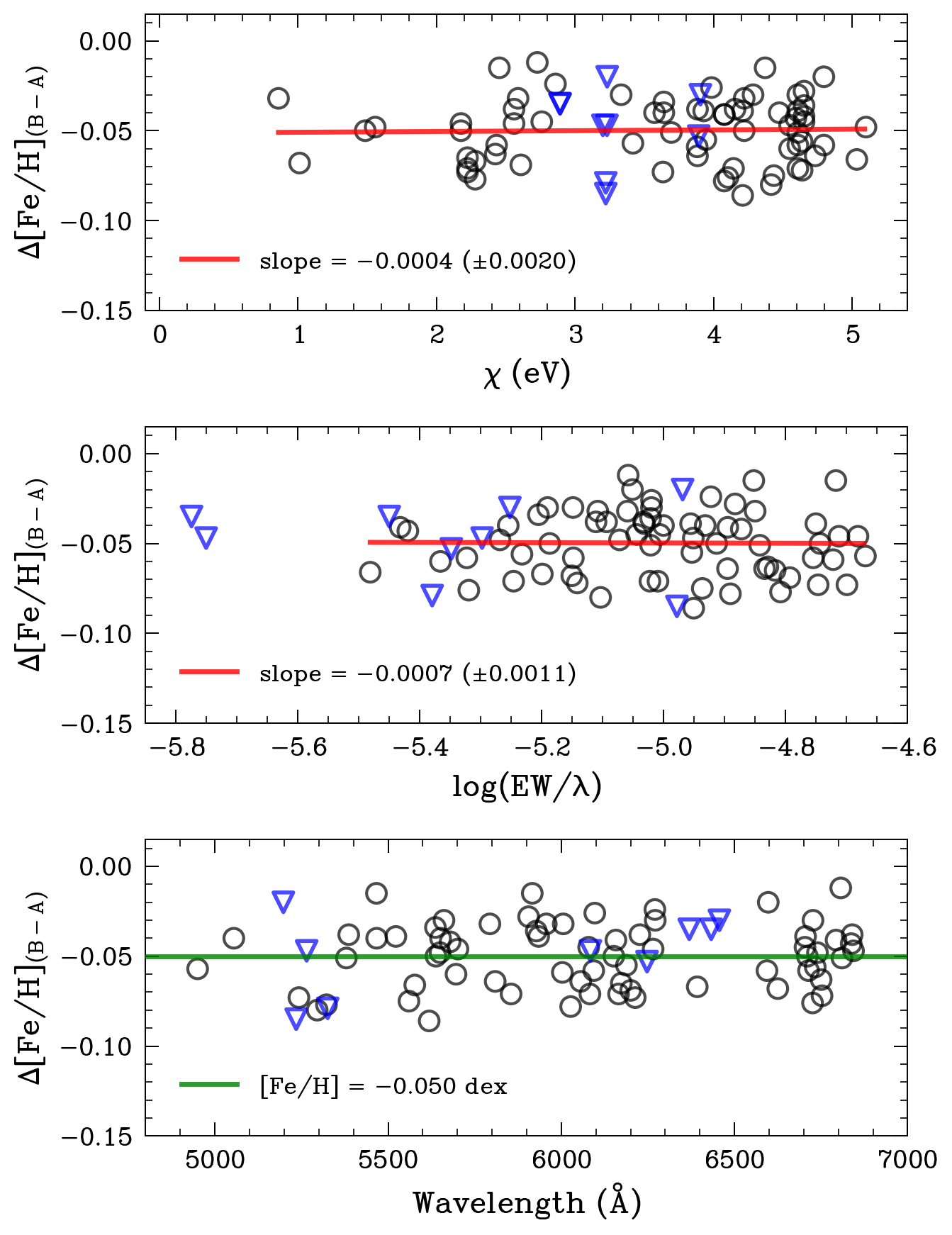}
 \centering
 \caption{Iron abundances measured in the spectrum of TOI-1173 B relative to TOI-1173 A as a function of excitation potential (top panel), reduced equivalent width (middle panel), and wavelength (bottom panel). Black open circles and blue triangles correspond to Fe I and Fe II lines, respectively. In the top and middle panels, red solid lines are linear fits to the Fe I abundances. In the lower panel, the green solid line is at the average $\Delta$[Fe/H] value. The adopted stellar parameters for the reference star TOI-1173 A were inferred using \textsc{xiru}.}
 \label{fig:SP_BA_xiru}
\end{figure}

\begin{figure}
 \includegraphics[width=\columnwidth]{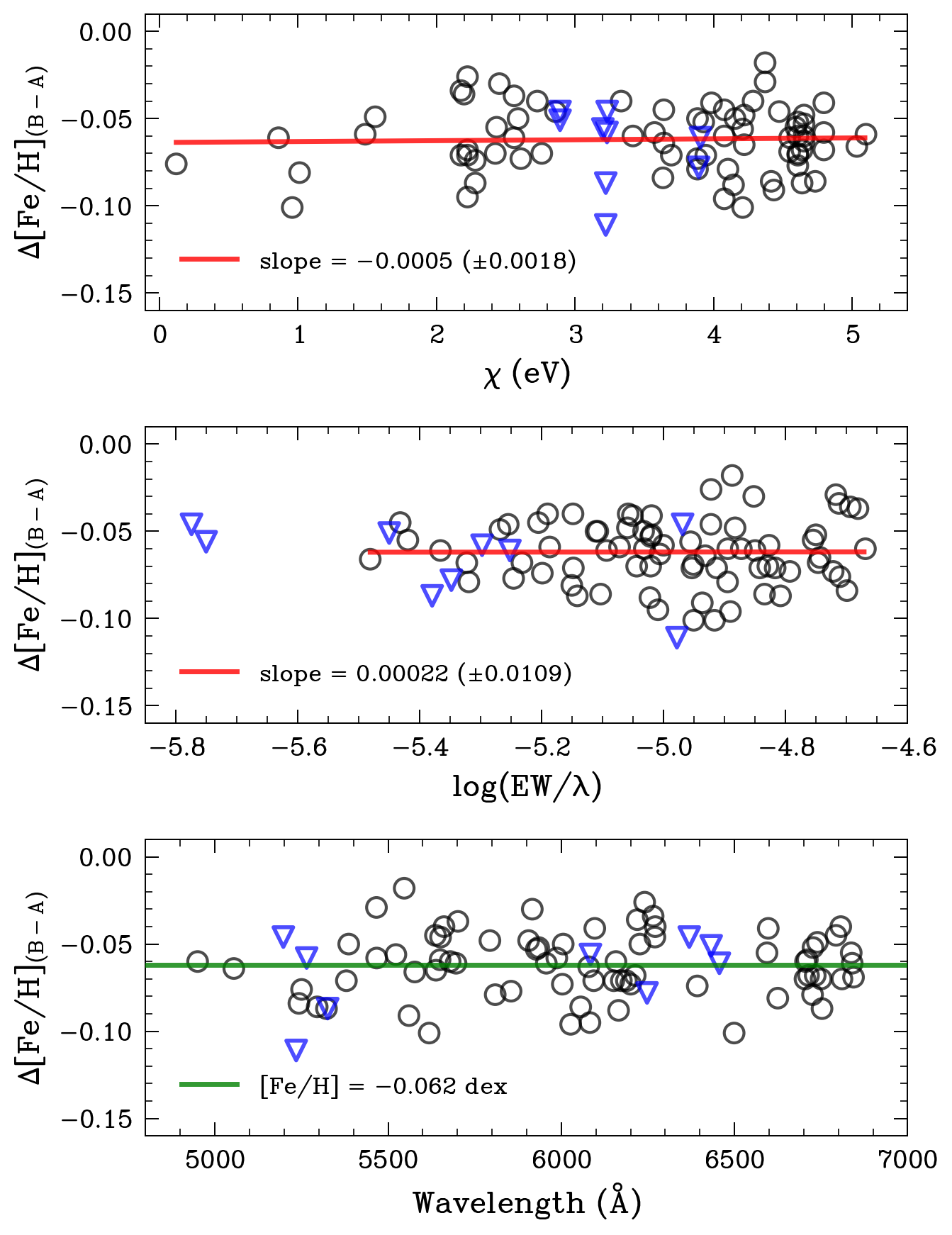}
 \centering
 \caption{As in Fig. \ref{fig:SP_BA_xiru}, but with the stellar parameters of the reference star TOI-1173 A inferred using \textsc{isochrones}.}
 \label{fig:SP_BA}
\end{figure}

Among the three methods discussed above, \textsc{xiru} and \textsc{isochrones} provide consistent \logg\ values with the trigonometric \logg. Therefore, we will discuss both methods in the following sections and how this affects the chemical abundance determinations of the pair. 

For the differential approach between components, which allows us to maximize precision in stellar parameters, we chose the A component as the reference star because it is brighter and has a higher SNR spectrum. We used \qq\ to determine the stellar parameters of component B relative to A (i.e., (B-A)), first using the stellar parameters of A estimated with \textsc{xiru}, and then employing \textsc{isochrones}. The results are listed in the last two rows of Table \ref{tab:fundamental_parameters}. Fig. \ref{fig:SP_BA_xiru} and \ref{fig:SP_BA} show that the spectroscopic equilibrium for the B component relative to A is satisfied.

\subsection{Ages, masses and radii}
\label{sec:ages}
As an output of the \textsc{Isochrones} fitting described above, we also computed stellar fundamental parameters (namely masses, ages, luminosities, radii). We obtained masses of $\textrm{M}=0.90^{+0.02}_{-0.02}$ \sm\ for star A and $\textrm{M}=0.83^{+0.02}_{-0.02}$ \sm\ for star B. The inferred ages were $\tau=9.3^{+1.4}_{-1.5}$ and $\tau=8.3^{+3.0}_{-3.0}$ Gyr, for components A and B, respectively. 

Alternatively, we derived masses, ages, and radii through another Bayesian approach anchored on a different set of structural models, Yonsei-Yale evolutionary tracks \citep{Yi:2001ApJS..136..417Y, Demarque:2004ApJS..155..667D}. This method is similar to the one employed in \cite{Grieves:2018MNRAS.481.3244G} and \cite{Diego:2019MNRAS.485L..68L}. In brief, considering the input parameters and its respective errors (\teff, \feh, [$\alpha$/Fe], \logg, \textit{Gaia} parallaxes, and $G$, $G_{\rm{BP}}$, and $G_{\rm{RP}}$ magnitudes), the posterior distribution functions were derived from a proper likelihood marginalization along all possible evolutionary steps. During this task, the likelihood was weighted by its respective mass (the \citealt{Salpeter:1955ApJ...121..161S} mass function) and metallicity \citep{Casagrande:2018IAUS..330..206C} priors. Finally, the desired evolutionary parameter's median (50th percentile) and the $\pm1 \sigma$ intervals (16th to 84th percentile) were obtained from their respective posterior cumulative distributions. The ages estimated for the A and B component are $\tau=8.7^{+2.0}_{-1.9}$ Gyr and $\tau=8.6^{+3.1}_{-3.5}$ Gyr, respectively. The mass inferred for star A is $0.911^{+0.028}_{-0.030}$ \sm\ and for star B is $0.839^{+0.033}_{-0.036}$ \sm.

\begin{table}
    \centering
	\caption{Ages, masses, and radii of the components of the binary system TOI-1173 A/B. All these parameters were inferred through a Bayesian approach that uses the Yonsei-Yale evolutionary models \citep{Yi:2001ApJS..136..417Y, Demarque:2004ApJS..155..667D}.}
    \begin{tabular}{lccr}
    \hline 
	\hline
    {\bf ID}                    &   Age (Gyr)       &  Mass (\sm)          &  Radius  ($R_{\odot}$)   \\
    \hline
	\multicolumn{4}{c}{Using \textsc{xiru}-based stellar parameters} \\
    \hline
	TOI-1173 A                  &  $8.3^{+1.9}_{-1.3}  $  & $0.910^{+0.013}_{-0.018} $ &  $0.928^{+0.013}_{-0.013} $ \\ 
	TOI-1173 B                  &  $8.1^{+2.5}_{-1.3}  $  & $0.838^{+0.013}_{-0.018} $ &  $0.805^{+0.010}_{-0.013} $ \\
	\hline
    \multicolumn{4}{c}{Using \textsc{isochrones}-based stellar parameters} \\
	\hline
    TOI-1173 A                  &  $8.7^{+2.0}_{-1.9}  $  & $0.911^{+0.028}_{-0.030} $ &  $0.934^{+0.011}_{-0.011} $ \\ 
	TOI-1173 B                  &  $8.6^{+3.1}_{-3.5}  $  & $0.839^{+0.033}_{-0.036} $ &  $0.820^{+0.010}_{-0.011} $ \\
    \hline
    \hline
    \end{tabular}
    \label{tab:age}
\end{table}

\begin{table*}
    \centering
	\caption{$^{(a)}$Differential chemical abundances of B relative to A inferred using {\sc xiru}-based stellar parameters. $^{(b)}$ Model engulfment for abundances from {\sc xiru}. $^{(c)}$ Differential chemical abundances of B relative to A estimated using {\sc isochrones}-based stellar parameters. $^{(d)}$ Model engulfment for abundances from {\sc isochrones}. The last column is the 50\% dust condensation temperature of elements \citep{Lodders:2003ApJ...591.1220L}. $^{(e)}$Non-LTE absolute abundances of lithium (\linlte) based on corrections of \citet{Lind:2009A&A...503..541L}. $^{(f)}$ Oxygen non-LTE corrected abundances using the grades of \citet{Ramirez:2007A&A...465..271R}. $^{(g)}$ Weighted average of \ion{Fe}{1} and \ion{Fe}{2}.}
	\begin{tabular}{lcccccr} 
		\hline
		\hline
		Element            &   Z   & TOI-1173 (B-A)$^{(a)}$     &    Model $^{(b)}$                   & TOI-1173 (B-A)$^{(c)}$     &     Model $^{(d)}$                   &  $T_{\rm{Cond}}$  \\   
		                   &       &    $\Delta$[X/H] (dex)     & $\Delta$[X/H]$_{\rm{(B-A)}}$ (dex)  &    $\Delta$[X/H] (dex)     & $\Delta$[X/H]$_{\rm{(B-A)}}$ (dex)   &       (K)         \\   
        \hline                                                                                                                                                                                              
		Li$^{(e)}$         &   3   &     0.010  $\pm$ 0.100     &       2.211                         &     0.010  $\pm$ 0.100     &       2.211                          &        1142       \\ 
		\ion{C}{1}         &   6   &  $-$0.021  $\pm$ 0.011     &    $-$0.001                         &  $-$0.030  $\pm$ 0.080     &    $-$0.001                          &        40         \\
        \ion{N}{1}         &   7   &     0.020  $\pm$ 0.090     &    $-$0.000                         &     0.090  $\pm$ 0.100     &    $-$0.000                          &        40         \\
        \ion{O}{1}$^{(f)}$ &   8   &     0.030  $\pm$ 0.030     &    $-$0.022                         &     0.032  $\pm$ 0.022     &    $-$0.020                          &        180        \\
        \ion{Na}{1}        &   11  &  $-$0.021  $\pm$ 0.010     &    $-$0.034                         &  $-$0.012  $\pm$ 0.012     &    $-$0.028                          &        958        \\
        \ion{Mg}{1}        &   12  &  $-$0.081  $\pm$ 0.006     &    $-$0.070                         &  $-$0.088  $\pm$ 0.006     &    $-$0.074                          &        1336       \\
        \ion{Al}{1}        &   13  &  $-$0.077  $\pm$ 0.012     &    $-$0.081                         &  $-$0.081  $\pm$ 0.014     &    $-$0.084                          &        1653       \\
        \ion{Si}{1}        &   14  &  $-$0.091  $\pm$ 0.013     &    $-$0.073                         &  $-$0.102  $\pm$ 0.012     &    $-$0.076                          &        1310       \\
        \ion{S}{1}         &   16  &     0.035  $\pm$ 0.040     &    $-$0.019                         &     0.020  $\pm$ 0.050     &    $-$0.011                          &        664        \\
        \ion{K}{1}         &   19  &  $-$0.114  $\pm$ 0.057     &    $-$0.027                         &  $-$0.100  $\pm$ 0.060     &    $-$0.022                          &        1006       \\
        \ion{Ca}{1}        &   20  &  $-$0.032  $\pm$ 0.014     &    $-$0.079                         &  $-$0.025  $\pm$ 0.056     &    $-$0.082                          &        1517       \\
        \ion{Sc}{1}        &   21  &  $-$0.034  $\pm$ 0.064     &    $-$0.065                         &  $-$0.021  $\pm$ 0.049     &    $-$0.067                          &        1659       \\
        \ion{Sc}{2}        &   21  &  $-$0.074  $\pm$ 0.011     &    $-$0.065                         &  $-$0.063  $\pm$ 0.011     &    $-$0.067                          &        1659       \\
        \ion{Ti}{1}        &   22  &  $-$0.067  $\pm$ 0.027     &    $-$0.067                         &  $-$0.058  $\pm$ 0.022     &    $-$0.070                          &        1582       \\
        \ion{Ti}{2}        &   22  &  $-$0.097  $\pm$ 0.017     &    $-$0.067                         &  $-$0.091  $\pm$ 0.018     &    $-$0.070                          &        1582       \\
        \ion{V}{1}         &   23  &  $-$0.031  $\pm$ 0.016     &    $-$0.089                         &  $-$0.023  $\pm$ 0.014     &    $-$0.092                          &        1429       \\
        \ion{Cr}{1}        &   24  &  $-$0.044  $\pm$ 0.014     &    $-$0.074                         &  $-$0.037  $\pm$ 0.020     &    $-$0.078                          &        1296       \\
        \ion{Cr}{2}        &   24  &     0.011  $\pm$ 0.013     &    $-$0.074                         &  $-$0.000  $\pm$ 0.030     &    $-$0.078                          &        1296       \\
        \ion{Mn}{1}        &   25  &     0.021  $\pm$ 0.041     &    $-$0.047                         &     0.023  $\pm$ 0.050     &    $-$0.045                          &        1158       \\
        Fe$^{(g)}$         &   26  &  $-$0.049  $\pm$ 0.010     &    $-$0.068                         &  $-$0.058  $\pm$ 0.010     &    $-$0.072                          &        1334       \\
        \ion{Co}{1}        &   27  &  $-$0.100  $\pm$ 0.010     &    $-$0.060                         &  $-$0.087  $\pm$ 0.009     &    $-$0.064                          &        1352       \\
        \ion{Ni}{1}        &   28  &  $-$0.018  $\pm$ 0.035     &    $-$0.069                         &  $-$0.023  $\pm$ 0.050     &    $-$0.073                          &        1353       \\
        \ion{Cu}{1}        &   29  &  $-$0.073  $\pm$ 0.026     &    $-$0.041                         &  $-$0.075  $\pm$ 0.025     &    $-$0.035                          &        1037       \\
        \ion{Zn}{1}        &   30  &     0.035  $\pm$ 0.037     &    $-$0.019                         &     0.030  $\pm$ 0.030     &    $-$0.011                          &        726        \\
        \ion{Rb}{1}        &   37  &  $-$0.025  $\pm$ 0.016     &    $-$0.028                         &  $-$0.020  $\pm$ 0.050     &    $-$0.021                          &        800        \\
		\ion{Y}{2}         &   39  &  $-$0.105  $\pm$ 0.027     &    $-$0.067                         &  $-$0.102  $\pm$ 0.027     &    $-$0.069                          &        1659       \\
        \ion{Ba}{2}        &   56  &  $-$0.121  $\pm$ 0.018     &    $-$0.064                         &  $-$0.116  $\pm$ 0.018     &    $-$0.066                          &        1455       \\
        \ion{Ce}{2}        &   58  &  $-$0.156  $\pm$ 0.025     &    $-$0.079                         &  $-$0.130  $\pm$ 0.024     &    $-$0.082                          &        1478       \\
        \ion{Nd}{2}        &   60  &  $-$0.111  $\pm$ 0.011     &    $-$0.083                         &  $-$0.093  $\pm$ 0.010     &    $-$0.086                          &        1602       \\
        \ion{Eu}{2}        &   63  &  $-$0.050  $\pm$ 0.031     &    $-$0.076                         &  $-$0.038  $\pm$ 0.026     &    $-$0.079                          &        1356       \\
		\hline
		\hline
	\end{tabular}
	\label{tab:chemical_abundances}
\end{table*}

Even though the two inferences use slightly different methodologies and different isochrone tracks were derived from different models, both methods provide consistent ages and masses within the errors. The ages of our pair confirm that they are truly coeval stars, i.e., they were formed at approximately at the same time, although with different masses. The adopted ages and masses used in this work are listed in Table \ref{tab:age}.

\section{Chemical Composition} \label{sec:chem}

\subsection{Abundance differences between components (B-A)}
The best laboratories to test the star-planet connection are binary systems, particularly when one component hosts an exoplanet while the other does not. If the components of planet-hosting binary systems are co-natal and co-eval, possible departures in the differential abundance between components could be attributed to planet signatures. The binary pair TOI 1173 A/B is then the perfect testbed for this hypothesis, as they are co-natal and co-eval, and to the best of our knowledge only TOI-1173 A hosts an exoplanet \citep[see][]{Yana_Galarza:2024AJ....168...91G}. Therefore, we performed a differential analysis on TOI 1173 A/B, using the A component as the reference star (B-A) since the two stars are more similar to each other than either is to the Sun. 

The chemical abundances were estimated from an EW analysis, using a procedure similar to that for \ion{Fe}{1} and \ion{Fe}{2}, involving fitting Gaussians and selecting local continua in the normalized spectra. This process was performed for component B with component A as the reference. We measured high-precision abundances for 19 elements (\ion{O}{1}, \ion{Na}{1}, \ion{Mg}{1}, \ion{Al}{1}, \ion{Si}{1}, \ion{K}{1}, \ion{Sc}{1}, \ion{Sc}{2}, \ion{Ti}{1}, \ion{Ti}{2}, \ion{V}{1}, \ion{Fe}{1}, \ion{Fe}{2}, \ion{Co}{1}, \ion{Cu}{1}, \ion{Zn}{1}, \ion{Rb}{1}, \ion{Y}{2}, \ion{Ba}{2}, \ion{Ce}{2}, \ion{Nd}{2}, and \ion{Eu}{2}). Our line list accounts for hyperfine structure and isotopic splitting for the following elements: \ion{Sc}{1}, \ion{Sc}{2}, \ion{V}{1}, \ion{Mn}{1}, \ion{Co}{1}, \ion{Cu}{1}, \ion{Y}{2}, \ion{Ba}{2}, and \ion{Eu}{2} from \cite{McWilliam:1998AJ....115.1640M,Prochaska2000a,Prochaska2000b,klose2002,Cohen:2003ApJ...588.1082C,Blackwell-Whitehead2005a,Blackwell-Whitehead2005b,Lawler2014}, and from the Kurucz\footnote{\url{http://kurucz.harvard.edu/linelists.html}} line lists.

We also computed abundances for \ion{C}{1}, \ion{N}{1}, \ion{S}{1}, \ion{Ca}{1}, \ion{Mn}{1}, \ion{Cr}{1}, \ion{Cr}{2} and \ion{Ni}{1}, using spectral synthesis with the line list generated from \textsc{linemake}\footnote{\url{https://github.com/vmplacco/linemake}} \citep{Placco:2021RNAAS...5...92P} updated with more recent laboratory data, when available. The abundance of lithium was also estimated using spectral synthesis and is described in subsection \ref{sub:litio}. The choice to employ spectral synthesis instead of equivalent width (EW) measurements relies on the fact that these lines are strongly blended, resulting in overestimated EW measurements. To measure nitrogen, we initially determined carbon abundance from the forbidden [\ion{C}{1}] line at 8727 \AA\ and the optical \ion{C}{1} 5052 and 5053 \AA \ lines. Subsequently, we employed the average carbon value to calculate the nitrogen abundance from the CN-dominated lines at 7111 and 7113 \AA. We validated this procedure using the solar spectrum. Our results for carbon (A (C) = 8.400 $\pm$ 0.046 dex) and nitrogen (A (N) = 7.83 $\pm$ 0.07 dex) in the Sun are in good agreement with those (8.46 and 7.83 dex) estimated by \citet{Asplund:2021A&A...653A.141A}. The same procedure was applied to determine \ion{C}{1} and nitrogen for the TOI-1173 A/B. The chemical makeup of B relative to A is listed in Table \ref{tab:chemical_abundances}.

\begin{figure*}
 $\begin{array}{c}
 \includegraphics[width=1.9\columnwidth]{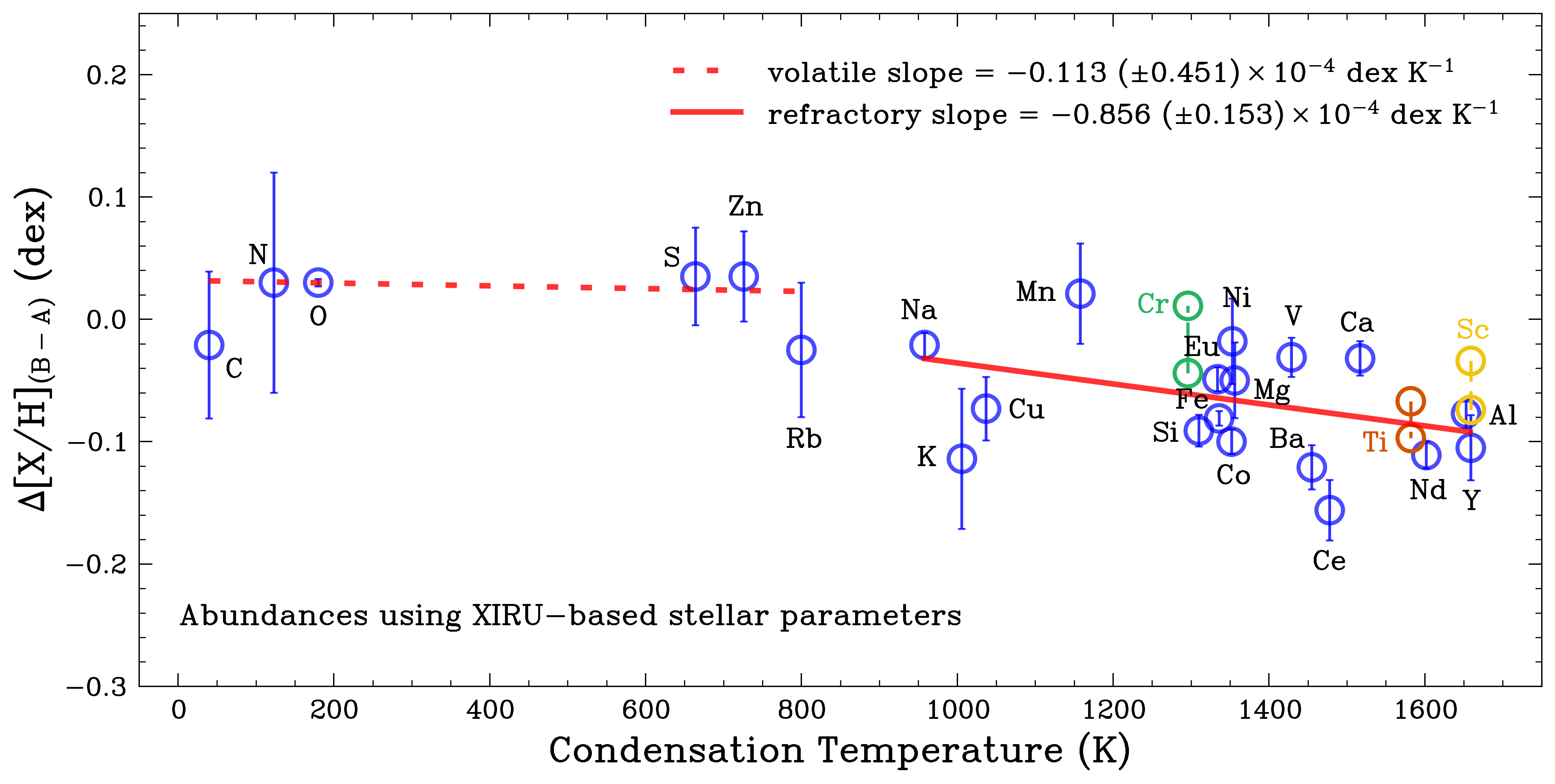} \\
 \includegraphics[width=1.9\columnwidth]{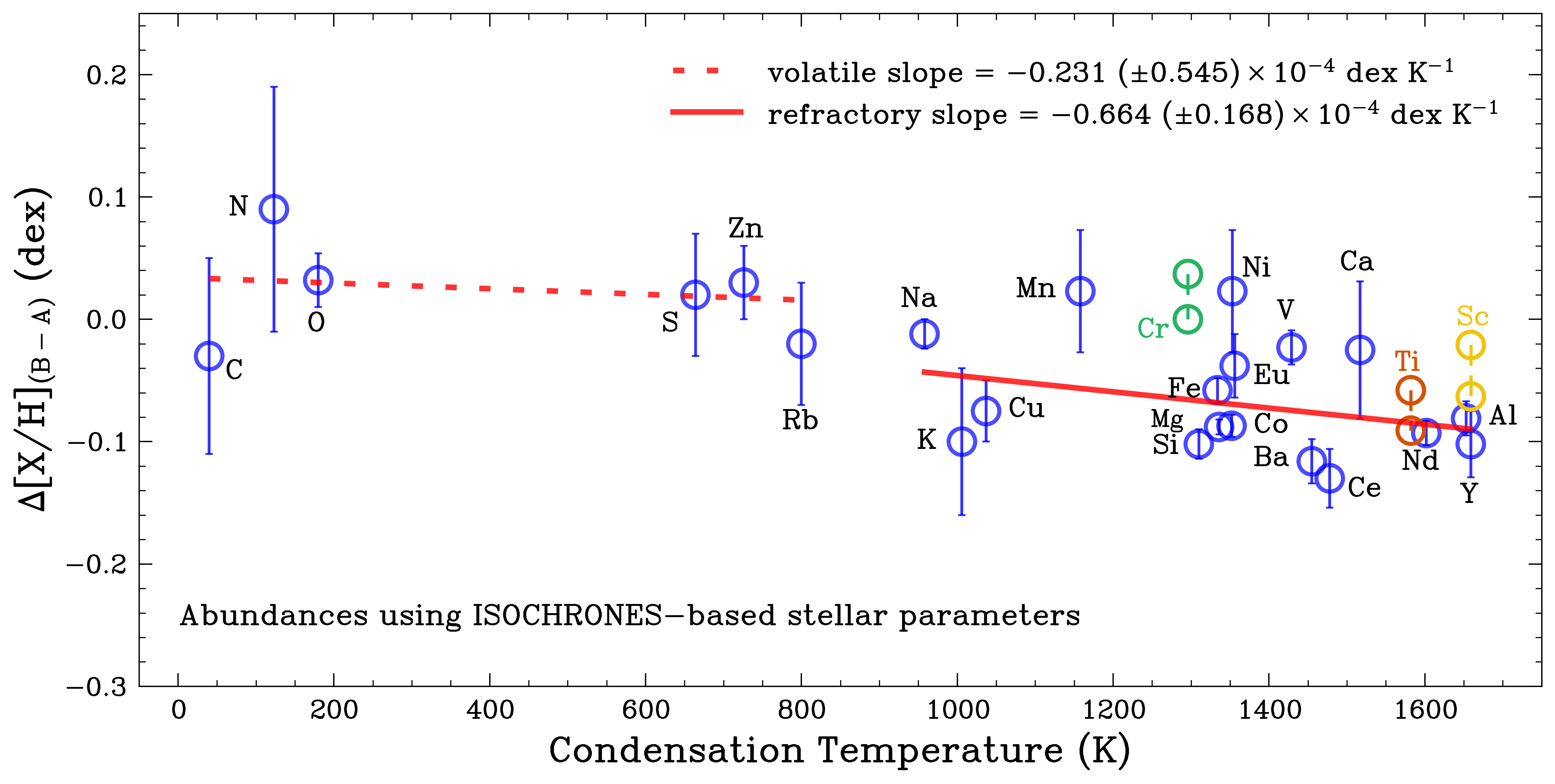}
 \end{array}$
 \centering
 \caption{Differential chemical abundances of TOI-1173 B relative to TOI-1173 A as a function of condensation temperature for stellar parameters estimated with {\sc xiru} (upper panel) and {\sc isochrones} (botton panel). The linear red solid and red dashed lines are the fits to the refractory (\Tc\ $>$ 900 K) and volatile elements (\Tc\ $<$ 900 K), respectively. Vertical dashed lines connect two species of the same chemical element (e.g., \ion{Cr}{1} and \ion{Cr}{2}).}
 \label{fig:tc_ab_BA}
\end{figure*}

Fig. \ref{fig:tc_ab_BA} shows the differential abundance between components (B-A) as a function of the condensation temperature for abundances inferred using the {\sc xiru}-based stellar parameters (upper panel) and {\sc isochrones}-based stellar parameters (bottom panel). Both panels display an abundance pattern in which component B is deficient in refractory elements compared to component A. However, both components display a similar abundance pattern in volatile elements. To determine whether the lower iron abundance for star B is due to uncertain differential atmosphere parameters relative to star A, we performed an LTE analysis using KURUCZ, ODFNEW, model atmospheres and MOOG. This was applied only to the abundances inferred with the {\sc isochrones} since the abundance pattern slope of this model is less significant. To increase the star B iron abundance (both \ion{Fe}{1} and \ion{Fe}{2}) by $\sim$0.05 dex, thus bringing it into agreement with star A, it was necessary to increase the differential $T_{\rm eff}$ of star B by 80 K and simultaneously increase \logg\ by 0.17 dex; this corresponds to 10$\sigma$ differential temperature change plus an 8$\sigma$ differential \logg\ change. Thus, it is not possible to make the iron abundance in star B and star A agree within the uncertainties of the differential atmosphere parameters, and we conclude that the deficit of Fe, and other refractory elements, in star B relative to star A is real.

It is worth noting that for (B-A), the abundance of Zn at 4722.159 \AA\ was estimated using ARCES spectra because the MAROON-X spectra presented artifacts at the Zn transition. Another challenging element to measure was \ion{K}{1}. The two strong potassium transitions (7664 and 7698 \AA) can be affected by ISM as well as telluric lines (in particular the $7664 \ $ \AA), and both transitions are typically highly saturated and with large non-LTE departures \citep{reggiani2019}. Other elements such as \ion{Na}{1}, \ion{V}{1}, \ion{Y}{2}, \ion{Ba}{2}, and particularly chromium, titanium and scandium, were also estimated using ARCES, and our results are in excellent agreement with our MAROON-X results. In the differential technique, departures from NLTE and GCE effects are minimized, and despite the fact that the components of the pair are not identical twins, the abundance pattern is evident. 

\begin{figure*}
  \begin{center}
  $\begin{array}{cc}
  \includegraphics[width=0.71\textwidth,angle=0]{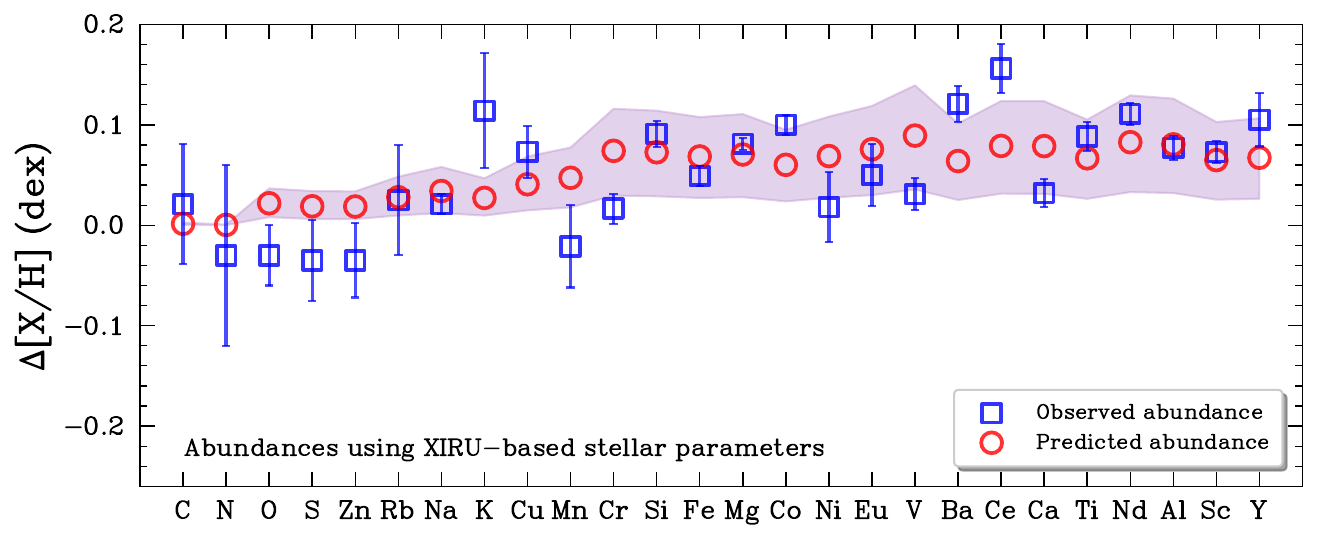}
  \includegraphics[width=0.28\textwidth,angle=0]{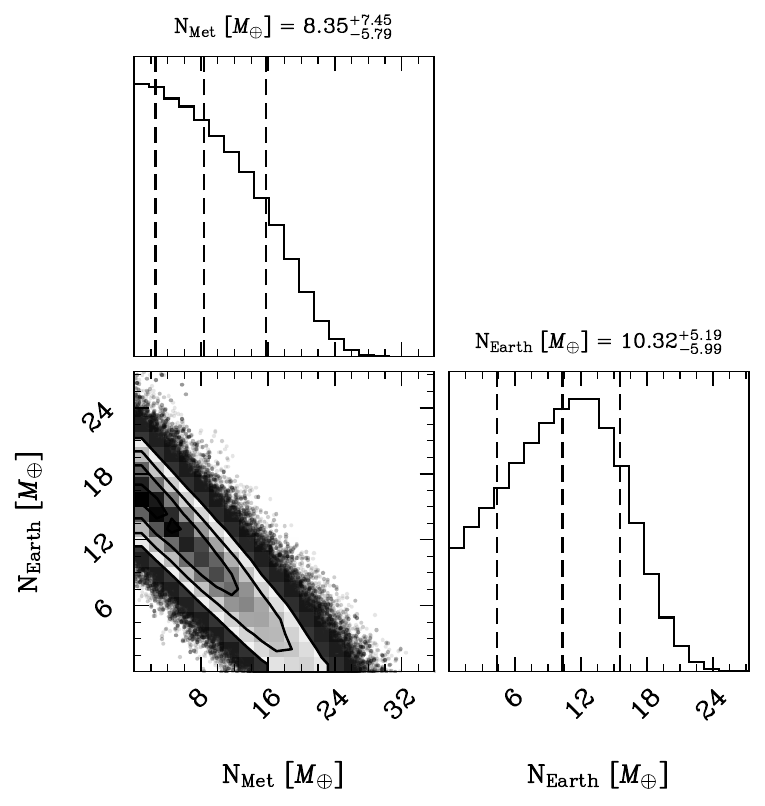} \\
  \includegraphics[width=0.71\textwidth,angle=0]{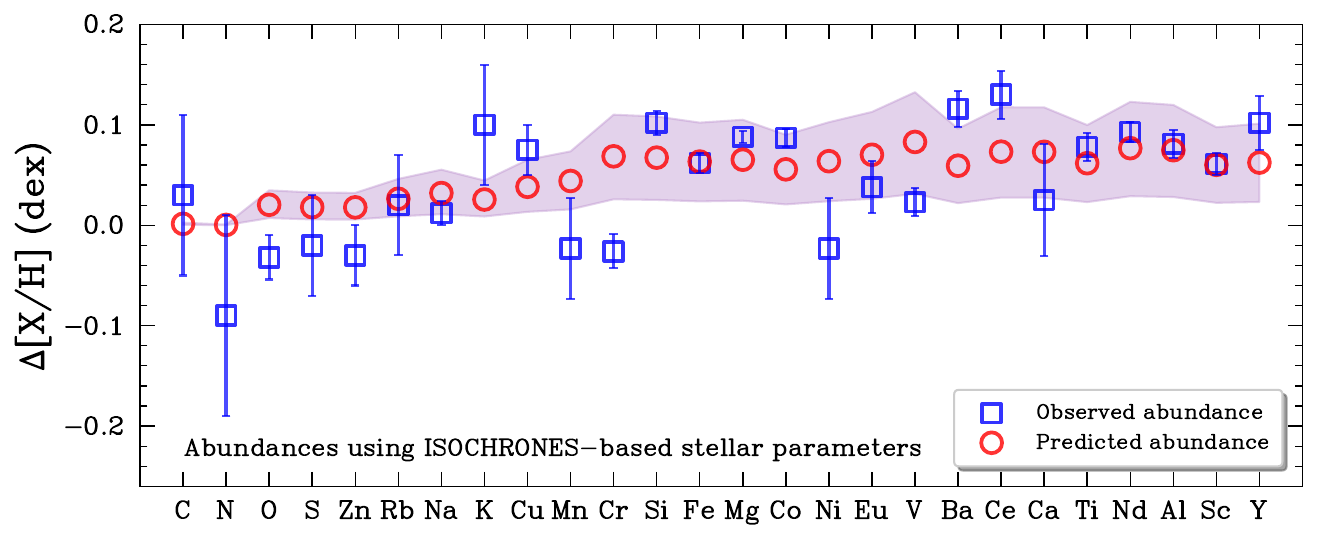}
  \includegraphics[width=0.28\textwidth,angle=0]{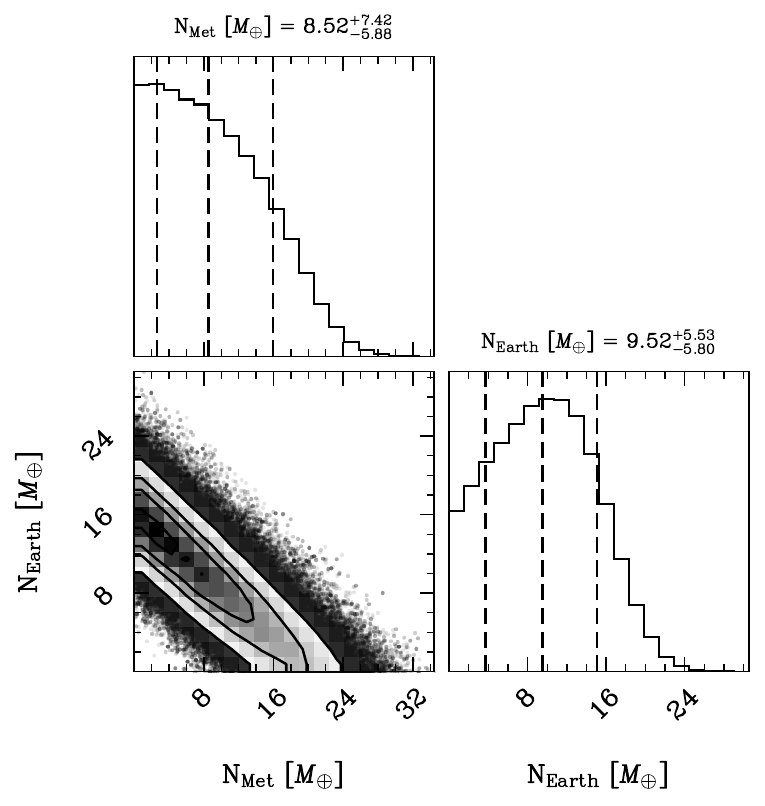}
  \end{array}$
  \caption{\textbf{Left panels:} Differential abundances of TOI-1173 A relative to B as a function of elements in ascending order of condensation temperature. The abundances are calculated using \textsc{xiru}-based (upper panel) and \textsc{isochrones}-based (bottom panel) stellar parameters. The observed and model abundances are represented by squares and circles, respectively. The shaded region represents the 1$\sigma$ posterior probability distribution. \textbf{Right panels:} Corner plot showing the results of our MCMC parameter estimation for our planet engulfment model. The dashed lines show the 16\% and 84\% percentiles for each parameter, thus representing 1$\sigma$ confidence ranges. The best-fit parameters are also indicated with black dashed lines.}
  \label{fig:engulfment_elements}
  \end{center}
\end{figure*}

\subsection{Lithium Abundance}
\label{sub:litio}
\citet{Carlos:2016A&A...587A.100C, Carlos:2019MNRAS.tmp..667C}, \citet{Martos:2023MNRAS.522.3217M} and \citet{Rathsam:2023MNRAS.525.4642R} have shown that the photospheric $^{7}$Li abundance (hereafter \li) in solar-type stars decreases over time. In order to determine \li\ in our pair, we synthesized the region around the asymmetric 6707.75 \AA\ Li line using the radiative transfer code \textsc{MOOG}, which assumes local thermodynamical equilibrium (LTE), with Kurucz model atmospheres. A detailed description of the method can be found in \citet{Yana_Galarza:2016A&A...589A..17Y} and \citet{Martos:2023MNRAS.522.3217M}. In both MAROON-X and ARCES spectra, we did not detect lithium, but we measured upper limits of \li$_{\rm{A}}$\ $=0.040\pm0.130$ and \li$_{\rm{B}}$\ $=0.010\pm0.130$. It can be observed that there is a consensus that both stars are severely depleted in lithium. In addition, the low lithium found in the pair is consistent with similar old ages, as older stars are more lithium depleted (e.g., see Fig. 2 in \citealp{Rathsam:2023MNRAS.525.4642R}). The \li\ upper limits were corrected for NLTE departures using the INSPECT database\footnote{\url{http://www.inspect-stars.com/}}, based on corrections of \citet{Lind:2009A&A...503..541L}. The obtained results are 0.150 dex for the A component and 0.160 dex for the B component. These values are in good agreement with the lithium-age anticorrelation reported by \citet{Martos:2023MNRAS.522.3217M}.

\section{Discussion} \label{sec:discussions}

\subsection{Planet formation and engulfment scenarios}
\label{sec:planet_scenarios}
Among the different hypotheses used to explain the chemical anomalies observed in solar-type stars and binary systems, especially in pairs exhibiting chemical anomalies between their components, three scenarios are more explored in the literature: planet engulfment, formation of rocky planets, and formation of giant gas planets. In this work, we will discuss these scenarios to explain the observed anomalies in our pair. 

During planet formation, both rocky and giant planets can deplete refractory elements in the host star (see Section \ref{sec:intro}). Since only TOI-1173 A hosts a Super-Neptune, we would expect to observe a depletion of refractory elements in the A component compared to B. However, the depletion is observed in the star without a known exoplanet, TOI-1173~B. One possible explanation could be that component B may host one or more massive exoplanets (total mass(es) larger than TOI-1173 A $b$), leading to the depletion of its refractory elements at a larger level. To date, we have not detected any transiting planets around TOI-1173~B, and the three MAROON-X spectra we obtained are not yet enough to detect exoplanet signatures via radial velocity variations.

In the engulfment scenario, an enhancement of refractory elements in the form of a positive \Tc-slope should be observed, followed by, in some cases, an enhancement of the lithium abundance. This scenario is best to explain the difference between planet-hosting star A relative to B, i.e., ($\textrm{A}-\textrm{B}$). To test this hypothesis, we used the \textsc{terra} code\footnote{\url{https://github.com/ramstojh/terra}}, which predicts the enhancement of volatile and refractory elements that one star may have as a result of the engulfment of a rocky exoplanet \citep[more details in][]{Yana_Galarza:2016A&A...589A..65G}. 

\begin{figure}
 $\begin{array}{c}
 \includegraphics[width=\columnwidth]{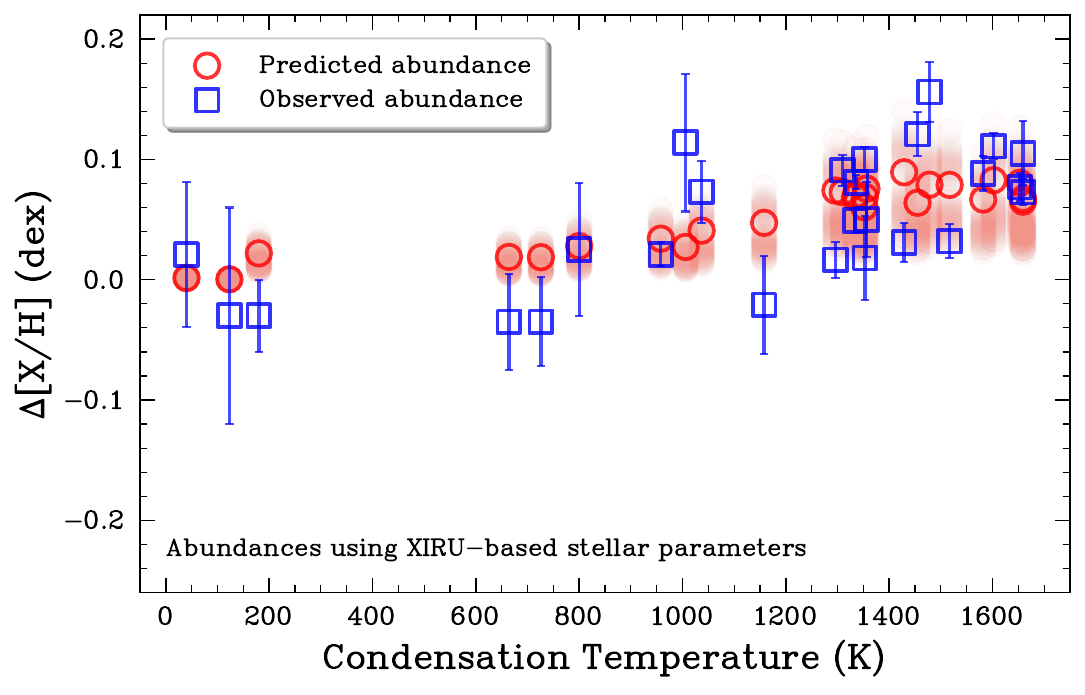} \\
 \includegraphics[width=\columnwidth]{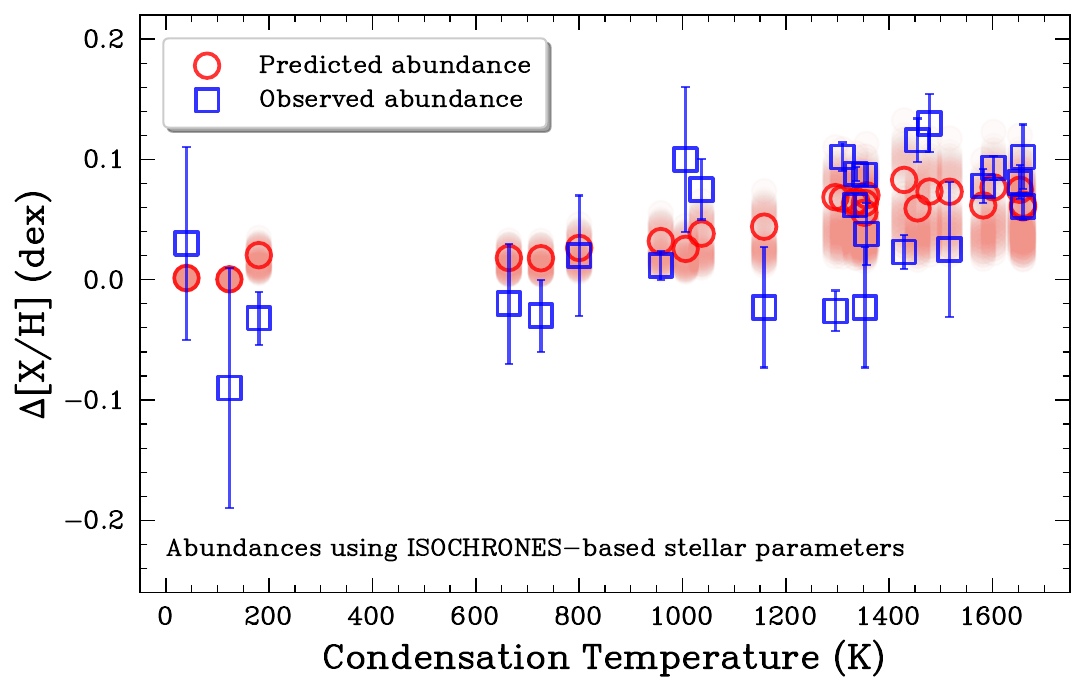}
 \end{array}$
 \centering
 \caption{Comparison of the abundances of TOI-1173 A relative to TOI-1173 B (depicted as blue open squares) as a function of dust condensation temperature for {\sc xiru}-based (upper panel) and {\sc isochrones}-based (bottom panel) stellar parameters. The predicted abundances are represented as red open circles and were estimated from a planetary engulfment of $\sim$18 M$_{\oplus}$ for both models. The blurred red circles show the $1\sigma$ posterior probability distribution.}
 \label{fig:engulfment}
\end{figure}

To reproduce the differential abundance pattern of ($\textrm{A}-\textrm{B}$), an engulfment of 18.67 M$_{\oplus}$ is required for the abundances estimated using {\sc xiru}-based stellar parameters. This entails a combination of $\approx$8.35$^{+7.45}_{-5.79}$ M$_{\oplus}$ of material with a chondritic composition and $\approx$10.32$^{+5.19}_{-5.99}$ M$_{\oplus}$ of Earth-like material (see upper panels of Fig. \ref{fig:engulfment_elements} and Fig. \ref{fig:engulfment}). For the abundance pattern inferred with {\sc isochrones}, the engulfment of 18.04 M$_{\oplus}$ (8.52$^{+7.42}_{-5.88}$ M$_{\oplus}$ plus $\approx$9.52$^{+5.53}_{-5.80}$ M$_{\oplus}$ of chondritic and Earth-like material, respectively, are necessary; see lower panels of Fig. \ref{fig:engulfment_elements} and Fig. \ref{fig:engulfment}). The hypothesis of ingestion of massive exoplanets appears plausible, considering the potential for the formation of such massive exoplanets, as demonstrated by \citet{Armstrong:2020Natur.583...39A}. These authors discovered the largest known terrestrial exoplanet, with a mass of 39.1 M$_{\oplus}$, which is likely the remnant core of a gas giant that lost its atmosphere through photoevaporation. Interestingly, the host star exhibits parameters similar to those of TOI-1173 A, implying that cores rich in refractory elements might preferentially form around cool, iron-rich stars. However, the detection of more massive terrestrial exoplanets is needed to confirm this.

The engulfment scenario predicts an approximately 2.2 dex enhancement in lithium abundance between components, contradicting the barely detected lithium in both components ($A$(Li)$_{\rm{(A-B)}} \sim 0.03$ dex). However, it is worth highlighting that \textsc{terra} assumes the engulfment occurred at the star's current age. Therefore, it is reasonable to speculate that the engulfment took place billions of years ago, allowing for the enhanced lithium to burn and reach the current low abundance. This is supported by the planet engulfment simulations of \citet{Sevilla:2022MNRAS.516.3354S}, which show that for stars with 0.9 \sm\ (similar to the mass of TOI-1173 A), the lithium abundance is quickly depleted after accretion, in timescales that are much shorter than depletion timescales for other chemical elements ($\sim1$ Gyr vs. $\sim10$ Gyr).

In \citet[][]{Yana_Galarza:2024AJ....168...91G}, we also investigated the dynamical timescales of TOI-1173 A in order to explore potential interactions between the primary star, its exoplanet, and the companion star, which could have led to the exoplanet's current orbit. Our results suggest that the timescales for both tidal circularization and von-Zeipel-Livov-Kozai (vZLK) perturbations range from $10^{8}- 10^{11}$ years, roughly corresponding to the age of the binary system. We therefore cannot rule out that the Super-Neptune TOI-1173 A $b$ might have experienced higher eccentricity in the past, which subsequently dampened, leading it to migrate to its current orbit over several billion years. 

\begin{figure*}
\includegraphics[width=2.1\columnwidth]{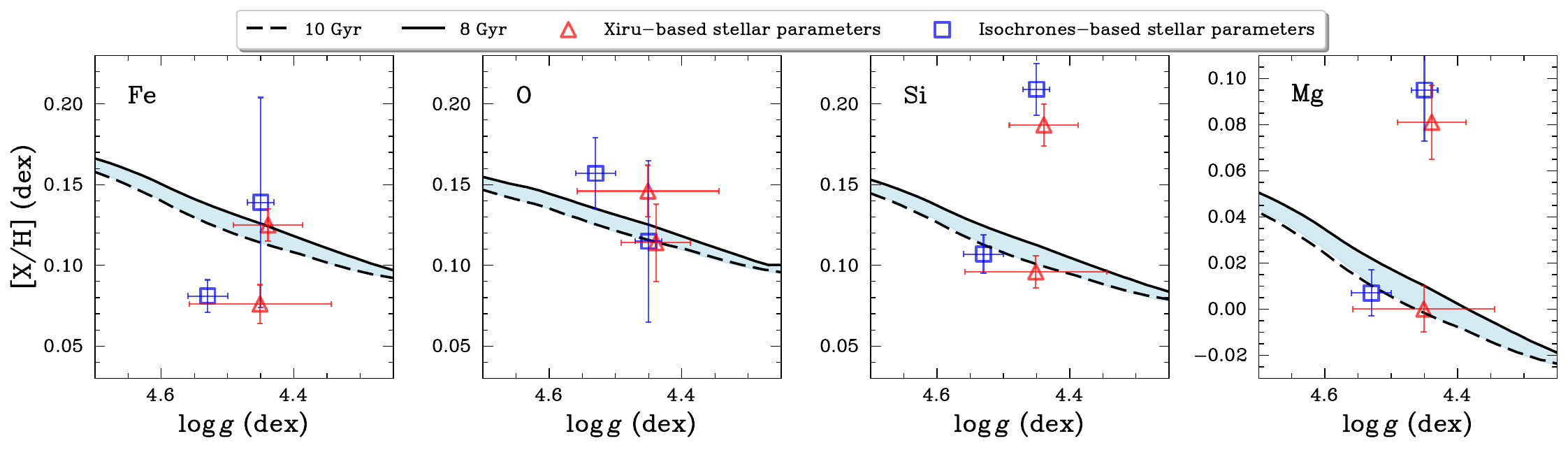}
 \centering
 \caption{Changing elemental abundances of iron, oxygen, silicon, and magnesium as a function of \logg. The blue shaded regions represent the predicted abundance changes for each element due to atomic diffusion, as modeled by \textsc{MIST} \citep{choi2016}. The models were computed for ages ranging from 8 (solid black lines) to 10 Gyr (dashed black lines). The red triangles and blue squares represent the observed abundances estimated using \textsc{xiru}-based and \textsc{isochrones}-based stellar parameters, respectively. Note that the abundance of some elements were vertically shifted to match the model, only to make it clear that atomic diffusion does not explain the abundance difference in the pair.}
 \label{fig:difussion}
\end{figure*}

Several studies have shown that close-in giant planets are formed through high-eccentricity tidal migration \citep[e.g.,][]{Rasio:1996Sci...274..954R, Wu:2003ApJ...589..605W, Fabrycky:2007ApJ...669.1298F, Raymond:2008MNRAS.384..663R, Chambers:2009AREPS..37..321C, Naoz:2012ApJ...754L..36N, Mustill:2015ApJ...808...14M, Church:2020MNRAS.491.2391C, Dong:2021ApJ...920L..16D, Rice:2023AJ....165...65R}. Tidal circularization, vZLK perturbations, or both may act as mechanisms to transport terrestrial planets inward. If a terrestrial planet reaches a semi-major axis of $a \lesssim 0.01$ AU, it will fall into the star \citep{Mardling:2004ApJ...614..955M, Raymond:2008MNRAS.384..663R}. Therefore, it is possible that the detected Super-Neptune orbiting star A might have initially formed with high eccentricity in the outer region of the protoplanetary disc and migrated inward over several billion years, reaching its current semi-major axis ($\sim0.07$ AU), which in turn could have driven interior terrestrial planets below the 0.01 AU limit, triggering their engulfment.

\subsection{Other scenarios} \label{sub:other_scenarios}
Given the projected separation, metallicity, and abundance patterns of TOI-1173 A/B, it is possible to think that the components are not associated and that they are a chance pair of unassociated single field stars moving together. \citet{Oh:2018ApJ...854..138O} addressed this possibility for the Kronos/Krios binary system using a distribution function of single stars in the Milky Way. They found that within 200 pc from the Sun, no stars have small differences between components such as $\Delta \mu_{\alpha}^{\ast} < 2$ mas yr$^{-1}$, $\Delta \mu_{\delta} < 2$ mas yr$^{-1}$, and $\Delta v_{r} < 2$ km s$^{-1}$, meaning that all stars within these values are not comoving by chance. TOI-1173 A/B is within this limit (see Table \ref{tab:parameters}), and along with the 3D velocity difference ($\Delta v_{3D} < 2$ km s$^{-1}$), reinforces the idea that both components are truly conatal.

Some works suggest that the chemical abundance differences observed in binary systems might be attributed to primordial inhomogeneities \citep[e.g.,][]{Ramirez:2019MNRAS.490.2448R, Fan_Liu:2021MNRAS.508.1227L, Behmard:2023MNRAS.521.2969B}. More recently, \citet{Saffe:2024A&A...682L..23S} determined the chemical composition of the giant-giant binary system HD 138202/CD-30 12303 and found a significant chemical difference between components at the level of 0.08 dex, which is unexpected as giant stars are sensitive to planet accretion and atomic diffusion effects. After testing those possibilities, they concluded that the chemical differences are attributed to primordial inhomogeneities as this does not produce condensation temperature trends. Following this, the abundance pattern observed in our pair rule out the scenario of primordial inhomogeneities.

Another possible scenario to explain the chemical difference in the pair is atomic diffusion. Microscopic diffusion and gravitational settling modify surface chemical abundances of main-sequence stars \citep[e.g.,][]{Michaud:1984ApJ...282..206M, dotter2016, Fan_Liu:2021MNRAS.508.1227L}. 
In particular, \citet{Fan_Liu:2021MNRAS.508.1227L} showed that binary systems with differences between components in \logg\ and \teff\ exceeding 0.07 dex and 200 K respectively, might exhibit non-negligible atomic diffusion effects. As the temperature difference in our pair is $\sim300$ K, we explored this possibility by generating stellar evolutionary models using the average metallicity of the pair and ages ranging from 8 to 10 billion years. We adopted the MIST \citep{Dotter:2016ApJS..222....8D} isochrones, based on MESA models, which include mixing via overshooting and atomic diffusion \citep{Choi:2016ApJ...823..102C}. Fig. \ref{fig:difussion} depicts the observed abundance difference for three refractory elements (iron, silicon and magnesium) and one volatile element (oxygen) as a function of surface gravity. The predicted change abundances due to atomic diffusion, estimated from \textsc{MIST}, are represented by the blue shaded region in each panel. It is clear that the observed abundance, represented by triangles  (\textsc{xiru}-based stellar parameters) and squares (\textsc{isochrones}-based stellar parameters), significantly exceeds the predicted model abundances, in particular for the refractory elements. Similar results are found for other elements. Therefore, we conclude that atomic diffusion does not fully explain the difference in chemical abundances observed between TOI-1173 A and B.

\section{Summary and Conclusions} \label{sec:conclusions}
We determined the precise stellar parameters and chemical composition of the planet-hosting wide binary TOI-1173 A/B using high-resolution MAROON-X and ARCES spectroscopy. To infer the stellar parameters, we employed two automatic codes ({\sc \qq} and {\sc xiru}) based on spectroscopic equilibrium, using the abundance of \ion{Fe}{1} and \ion{Fe}{2} to converge to the best solution, with the Sun as the reference star. We also use {\sc isochrones} to infer stellar parameters by fitting stellar models. Ages, masses, and radii were inferred via isochrone fitting using Bayesian methods. The resulting fundamental parameters are listed in Table \ref{tab:fundamental_parameters} and \ref{tab:age}. Our results indicate that both components are cool and iron-rich stars. The planet-hosting component (TOI-1173 A) is a G9-type star, while component B is a K1.5-type star according to \citet{Gray:2009ssc..book.....G}.

We obtained precise chemical abundances for 27 elements, using the Sun as the reference star, denoted as (A-Sun) and (B-Sun). The procedure is explained in Appendix \ref{appendix}. Figure \ref{fig:gce_thin_disk} shows the evolution of the elemental abundance of the stars within 150 pc in the Galactic disc. Both components of the pair follow the GCE trend observed in G-type stars. To compare the components with the Sun, we performed GCE corrections based on fits of elemental abundances and ages. The GCE-corrected abundances of both components were found to resemble those of the Sun in the condensation temperature plane for both volatile and refractory elements (see Fig. \ref{fig:AB_pattern_Sun_xiru} and \ref{fig:AB_pattern_Sun}). 

To minimize NLTE and other effects, we then employed the differential technique between components, taking the A component as the reference star. In Fig. \ref{fig:tc_ab_BA}, the differential abundance (B-A) is depicted as a function of condensation temperature, revealing that the B component is depleted in refractory elements. This implies that the component harboring the Super-Neptune is abundant in refractory elements, contradicting the planet formation hypothesis. We explored scenarios, such as charge captures, primordial inhomogeneities, and atomic diffusion (see subsection \ref{sub:other_scenarios}). None of these explain the observed abundance difference in the pair, except for atomic diffusion, which has a small contribution. The most promising scenario to explain the excess of refractory elements observed in star A is the engulfment of a rocky planet (or planets) of $\sim$18 M$_{\oplus}$ (see Fig. \ref{fig:engulfment_elements} and \ref{fig:engulfment}).

To investigate the planet engulfment hypothesis, we explored the dynamical timescales of TOI-1173 A and found that the Super-Neptune might have arrived at its observed short period through high-eccentricity migration. The potential mechanisms for the migration are tidal circularization and the von-Zeipel-Livov-Kozai perturbation, as their timescales are comparable to the age of the binary system. Therefore, it is possible that the migration of the Super-Neptune TOI-1173 A $b$ could have led to the engulfment of one or more massive rocky planets, enhancing the refractory elements in TOI-1173 A. This scenario might elucidate the observed chemical anomalies of (A-B) in the condensation temperature plane.

\section*{Acknowledgments}
Jhon Yana Galarza acknowledges support from their Carnegie Fellowship. Henrique Reggiani acknowledges the support from NOIRLab, which is managed by the Association of Universities for Research in Astronomy (AURA) under a cooperative agreement with the National Science Foundation. Thiago Ferreira acknowledges support from Yale Graduate School of Arts and Sciences. Diego Lorenzo-Oliveira acknowledges the support from CNPq (PCI 301612/2024-2).

This work made use of data collected with the Gemini Telescope. This paper includes data collected by the TESS mission. Funding for the TESS mission is provided by the NASA's Science Mission Directorate.  This work has made use of data from the European Space Agency (ESA) mission {\it Gaia} (\url{https://www.cosmos.esa.int/gaia}), processed by the {\it Gaia} Data Processing and Analysis Consortium (DPAC, \url{https://www.cosmos.esa.int/web/gaia/dpac/consortium}). Funding for the DPAC has been provided by national institutions, in particular the institutions participating in the {\it Gaia} Multilateral Agreement. This publication makes use of data products from the Two Micron All Sky Survey, which is a joint project of the University of Massachusetts and the Infrared Processing and Analysis Center/California Institute of Technology, funded by the National Aeronautics and Space Administration and the National Science Foundation.  This research has made use of the SIMBAD database, operated at CDS, Strasbourg, France \citep{Wenger:2000A&AS..143....9W}.  This research has made use of the VizieR catalogue access tool, CDS, Strasbourg, France (DOI: 10.26093/cds/vizier). The original description of the VizieR service was published in 2000, A\&AS 143, 23 \citep{Ochsenbein:2000A&AS..143...23O}.

\vspace{5mm}
\facilities{Gemini:Gillett, ARC, \textit{Gaia}, TESS}

\software{
\textsc{numpy} \citep{van_der_Walt:2011CSE....13b..22V}, 
\textsc{matplotlib} \citep{Hunter:4160265}, 
\textsc{pandas} \citep{mckinney-proc-scipy-2010}, 
\textsc{Astropy} \citep{astropy:2022},
\textsc{astroquery} \citep{Ginsburg:2019AJ....157...98G}, 
\textsc{iraf} \citep{Tody:1986SPIE..627..733T}, 
\textsc{iSpec} \citep{Blanco:2014A&A...569A.111B, Blanco:2019MNRAS.486.2075B}, 
\textsc{Kapteyn} Package \citep{KapteynPackage}, 
\textsc{moog} \citep{Sneden:1973PhDT.......180S}, 
q$^{\textsc{2}}$ \citep{Ramirez:2014A&A...572A..48R}, 
\textsc{Gala} \citep{gala}, 
\textsc{terra} \citep{Yana_Galarza:2016A&A...589A..65G}, 
\textsc{DYNESTY} \citep{2020MNRAS.493.3132S}, 
\textsc{astrobase} \citep{2021zndo...4445344B}, 
\textsc{xiru} \citep{Alencastro_thesis:1885-288051}.
\textsc{VOSA} \citep{2008A&A...492..277B}, 
\textsc{smplotlib} 
\citep{jiaxuan_li_2023_8126529}.
}

\appendix

\section{Abundance differences relative to the Sun}
\label{appendix}

Section \ref{sec:chem} describes in detail the calculation of the chemical abundance of the pair using the A component as the reference star. In this section, we apply the same procedure, but we compare each component of the binary system with the Sun. This has some disadvantages because the A and B components have significantly different temperatures and ages than the Sun. Therefore, GCE corrections and atomic diffusion must be considered. However, in subsection \ref{sub:other_scenarios}, it was demonstrated that atomic diffusion does not explain the chemical differences observed in the pair. GCE and non-LTE corrections are described below.


The chemical abundances for the pair relative to the Sun, meaning (A-Sun) and (B-Sun), are summarized in Table \ref{tab:chemical_abundances_sun}. The [$\alpha$/Fe]\footnote{[$\alpha$/Fe] = 1/4 ([Mg/Fe] + [Si/Fe] + [Ca/Fe]+ [Ti/Fe])}, [Mg/Fe] and [Ti/Fe] values for stars A and B are (0.021, 0.041, 0.012) and (0.003, 0.013, 0.00), respectively, using abundances from \textsc{isochrones}-based stellar parameters. These values are consistent with the kinematic thin-disc membership of the pair (see Fig. \ref{fig:Toomre_diagram}, and also Fig. 7 in \citealp{Adibekyan:2012A&A...545A..32A}, Fig. 3 in \citealp{Fuhrmann:2017MNRAS.464.2610F}, and Fig. 23 in \citealp{Bensby:2014A&A...562A..71B}). Similar results are found when using abundances with \textsc{xiru}-based stellar parameters.

\begin{figure*}
 \includegraphics[width=2.1\columnwidth]{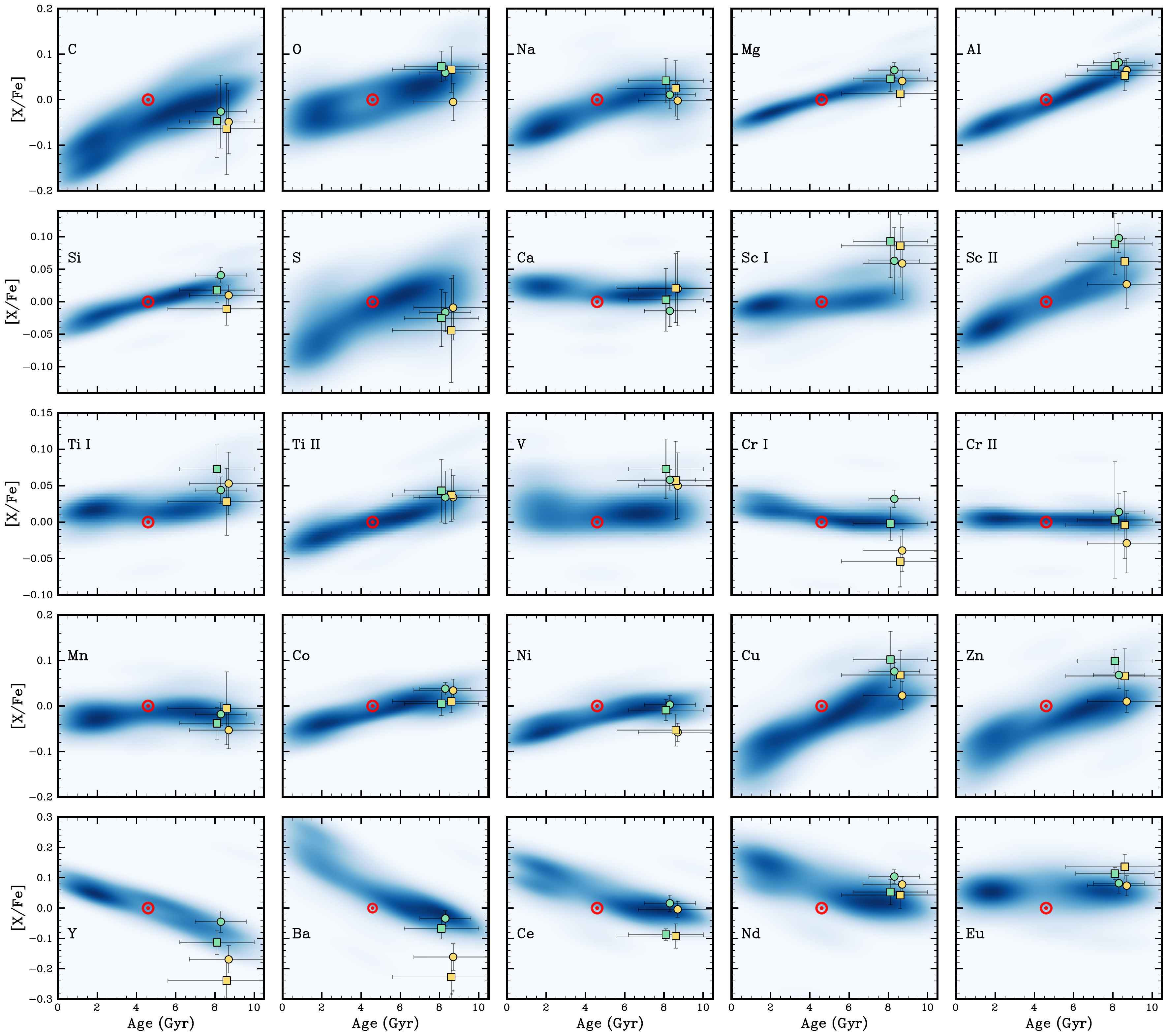}
 \centering
 \caption{Density plot of elemental abundances of G-type stars \citep{Bedell:2018ApJ...865...68B, Spina:2018MNRAS.474.2580S, Yana_Galarza:2021MNRAS.504.1873Y} in the thin-disk with 5600 K $\leq$ \teff\ $\leq 6000$ K as a function of age. The circle and square symbols represent TOI-1173 A and TOI-1173 B, respectively. The green symbols indicate abundances using \textsc{xiru}-based stellar parameters, while the yellow symbols use \textsc{isochrones}-based stellar parameters. For oxygen, the yellow and green symbols represent the NLTE-corrected abundance estimated using the grids of \citet{Ramirez:2007A&A...465..271R}. The Sun is plotted as a red solar standard symbol. The ages of the G-type stars were estimated using the Yonsei-Yale evolutionary tracks \citep{Yi:2001ApJS..136..417Y}, the same models used to infer the adopted age of TOI-1173 A/B.}
 \label{fig:gce_thin_disk}
\end{figure*}

Fig. \ref{fig:gce_thin_disk} shows a density plot of an updated Galactic Chemical Evolution (GCE) pattern for 25 elements in thin-disc G stars. The new GCE pattern is based on the solar twin sample of \citet{Bedell:2018ApJ...865...68B}, but updated with the Inti sample of solar-type stars of \citet{Yana_Galarza:2021MNRAS.504.1873Y}. The circles and squares represent the chemical composition of components A and B, while the yellow and green colors indicate the abundances using \textsc{isochrones}-based and \textsc{xiru}-based stellar parameters. It is important to mention that, as the \ion{O}{1} abundances of the new sample of G stars were corrected for non-LTE effects using the grids of \citet{Ramirez:2007A&A...465..271R}, for consistency, we applied the same non-LTE corrections to our pair. Thus, the green symbols in the oxygen trend represent the NLTE-correction from this study. From Fig. \ref{fig:gce_thin_disk}, we see that the chemical abundance of both components follows the typical GCE trend of G stars, including the NLTE corrected abundances. Nonetheless, both components display departures in the \ion{C}{1}, \ion{Y}{2}, and \ion{Ba}{2} elements relative to the GCE trends for G stars when using \textsc{isochrones}-based stellar parameters. We investigated the possible systematics due to the fact that only one line was available for \ion{Y}{2} in MAROON-X spectra by calculating the same abundances from the ARCES spectra, and found consistent results. Similar results were found for \ion{Ba}{2} using ARCES. Thus, we conclude that the origin of the departures is solely due to the choice of method, as the abundances with \textsc{xiru}-based stellar parameters better follow the GCE trend. Interestingly, the B component, a K-type star, follows the GCE trend of G-type stars, indicating that we can apply our GCE corrections (see below).

In order to accurately compare the pair with the Sun, it is necessary to consider the GCE of stars. Comprehensive studies such as those of \cite{Nissen:2015A&A...579A..52N, Bedell:2018ApJ...865...68B} and \cite{Spina:2018MNRAS.474.2580S} have provided crucial insights into GCE in the solar neighborhood. These studies established GCE trends for solar twin stars by fitting individual abundances and ages, while also accounting for \feh\ effects. In this study, we use our updated GCE corrections, which are based on data from 200 G-type stars (Yana Galarza et al., 2024b, in prep.). For all the elements, we fitted the [Fe/H] versus age relation using linear and polynomial models that account for the uncertainties in [Fe/H] and ages. Once we found the parameters that define the best-fitting line and curve, we use these models to correct the abundances of the pair for GCE effects. Since the system is older than the Sun, we removed the GCE effects. We refer to \citet{Spina:2016A&A...593A.125S, Yana_Galarza:2016A&A...589A..17Y, Bedell:2014ApJ...795...23B} for more details.

\begin{figure}
 \includegraphics[width=\columnwidth]{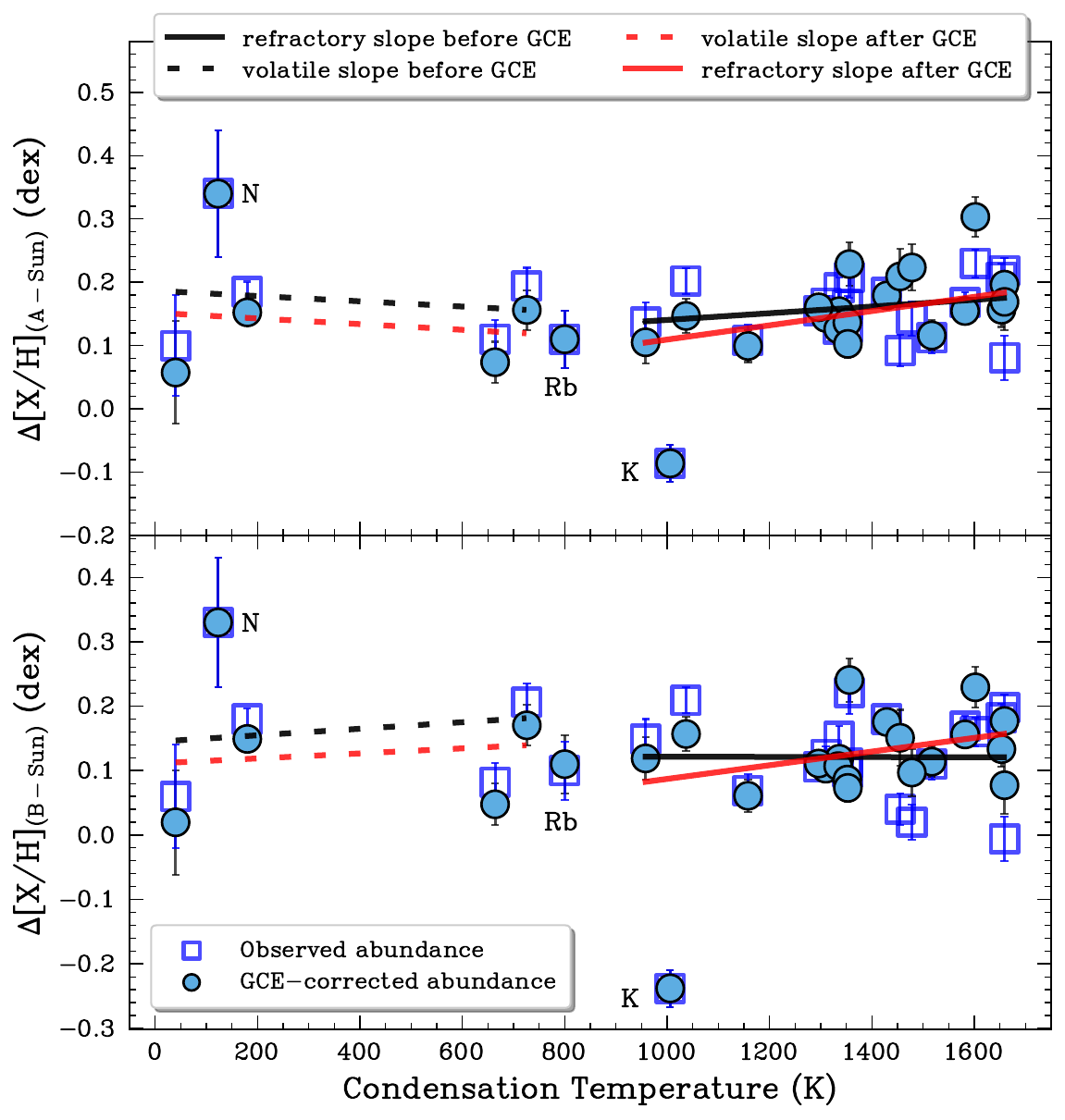}
 \centering
 \caption{Observed (open squares) and GCE-corrected (filled circles) differential abundances of TOI-1173 A (upper panel) and TOI-1173 B (bottom panel) relative to the Sun as a function of dust condensation temperature. The abundances were estimated using \textsc{xiru}-based stellar parameters. The green and red solid lines represent the fits to refractory elements (\Tc\ $>$ 900 K) before and after applying GCE corrections. The same applies to the dashed lines representing volatile elements (\Tc\ $<$ 900 K). The elements \ion{N}{1}, \ion{Rb}{1}, and \ion{K}{1} are not included in the fits.}
 \label{fig:AB_pattern_Sun_xiru}
\end{figure}

\begin{figure}
 \includegraphics[width=\columnwidth]{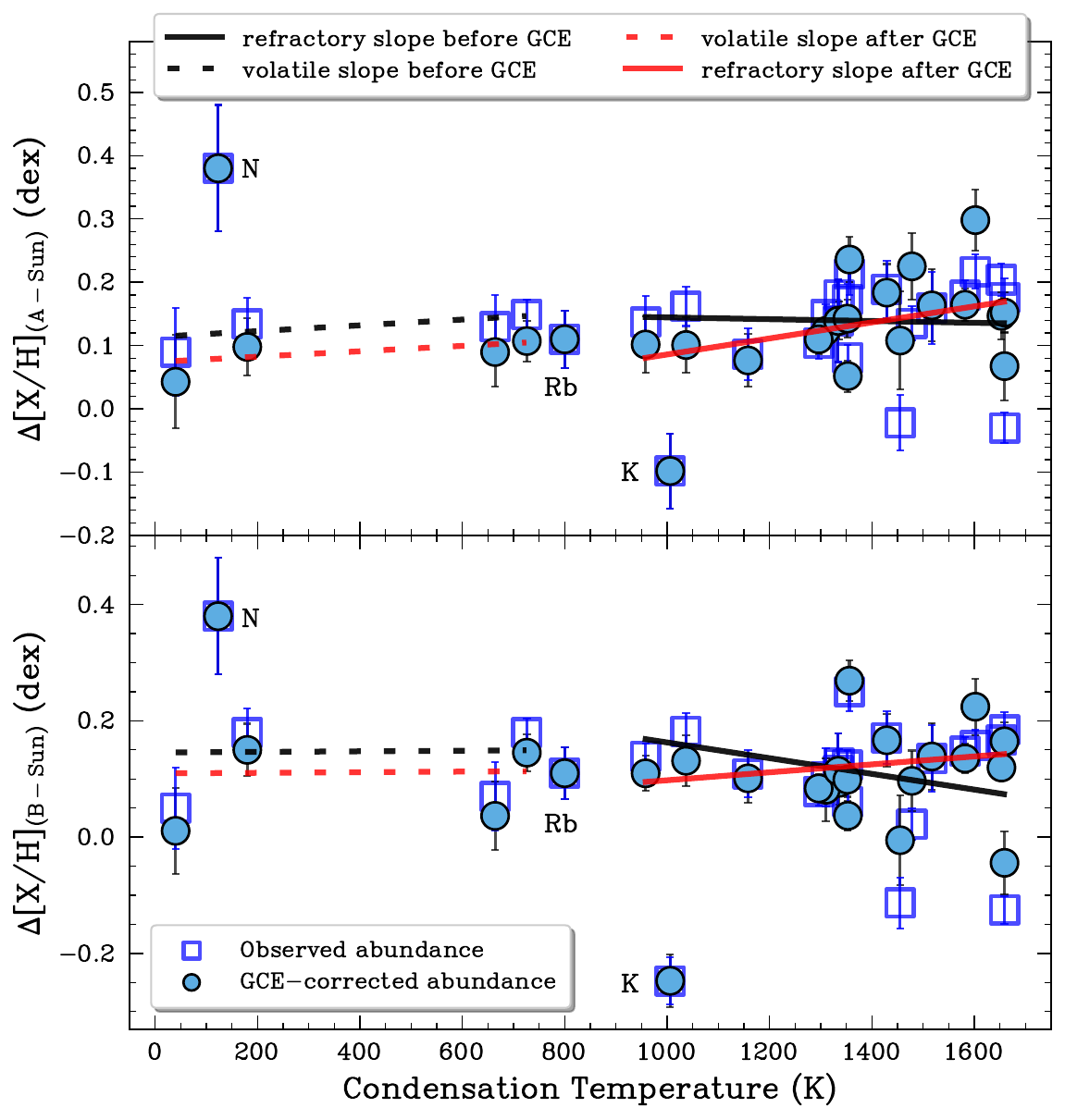}
 \centering
 \caption{Same as in Fig. \ref{fig:AB_pattern_Sun_xiru}, but with abundances estimated using \textsc{isochrones}-based stellar parameters.}
 \label{fig:AB_pattern_Sun}
\end{figure}

In Fig. \ref{fig:AB_pattern_Sun_xiru} (\textsc{xiru}-based stellar parameters) and \ref{fig:AB_pattern_Sun} (\textsc{isochrones}-based stellar parameters), the differential abundance of star A (upper panel) and B (bottom panel) relative to the Sun in the condensation temperature \citep[taken from the 50\% condensation temperature values inferred by][]{Lodders:2003ApJ...591.1220L} plane is shown. We consider the volatile elements to have \Tc\ $<$ 900 K, while refractory elements have \Tc\ $>$ 900 K, as stated in \citet{Melendez:2009ApJ...704L..66M}. 

The linear fits were carried out using the Fitter object of the \textsc{KAPTEYN} kmpfit package, which takes into account abundance uncertainties. The slopes of the fits for the refractory elements in components A and B before and after GCE correction are listed in Table \ref{tab:slopes}.

\begin{table}
    \centering
	\caption{\textbf{}}
    \begin{tabular}{lcr}
    \hline 
	\hline
    Category       &   Slope (dex K$^{-1}$)          &  {Method}                        \\
    \hline
    \multicolumn{3}{c}{Before GCE corrections}                                                      \\
    \hline
    Refractory (A-Sun)   &  $0.529 (\pm0.265)\times 10^{-4}$     & \textsc{xiru}                    \\ 
    Refractory (B-Sun)   &  $-0.013 (\pm0.440)\times 10^{-4}$    & \textsc{xiru}                    \\ 
	Refractory (A-Sun)   &  $-0.142 (\pm0.359)\times 10^{-4}$    & \textsc{isochrones}              \\
    Refractory (B-Sun)   &  $-1.351 (\pm0.469)\times 10^{-4}$    &  \textsc{isochrones}             \\ 
    \hline
    \multicolumn{3}{c}{After GCE corrections}                                                       \\
    \hline
    Refractory (A-Sun)   &   $1.125 (\pm0.303)\times 10^{-4}$   & \textsc{xiru}                    \\ 
    Refractory (B-Sun)   &   $1.061 (\pm0.512)\times 10^{-4}$    & \textsc{xiru}                    \\ 
	Refractory (A-Sun)   &   $1.267 (\pm0.466)\times 10^{-4}$    & \textsc{isochrones}              \\ 
    Refractory (B-Sun)   &   $0.682 (\pm0.613)\times 10^{-4}$    & \textsc{isochrones}              \\
    \hline
    \hline
    \end{tabular}
    \label{tab:slopes}
\end{table}

From Fig. \ref{fig:AB_pattern_Sun_xiru} and \ref{fig:AB_pattern_Sun}, it is clear that the \Tc-slopes for volatile elements, both before and after GCE-corrections, are not statistically significant, indicating that both components exhibit similar abundance patterns to the Sun. The only statistically significant \Tc-slope observed for the refractory elements is the A component after GCE-corrections, although at a marginal significance level of 2.7$\sigma$. Thus, we conclude that both components exhibit abundance patterns similar to the Sun after being corrected for GCE-effects. It is worth highlighting that \ion{N}{1} and \ion{K}{1} elements are excluded from our fits because they are anomalous compared to the Sun and do not follow the volatile/refractory abundance trend observed in both stars. Additionally, although \ion{Rb}{1} follows the volatile trend for our pair, we exclude it from our fits, as no GCE correction is available. 

Discrepancies of some elements are the typical challenge faced in differential analysis when the target and the reference star do not have similar enough stellar parameters. In that case, uncertainties arising from atmospheric models and NLTE effects are not minimized, thus affecting the precision of the comparison and potentially obscuring signatures of planet formation or engulfment. Here, a differential analysis between the components is the most appropriate approach given the modest temperature difference between the two stars. 

\begin{table*}
    \centering
	\caption{$^{(a)}$Differential abundances relative to the Sun using \textsc{xiru}-based stellar parameters. $^{(b)}$Oxygen non-LTE corrected abundances using the grades of \citet{Ramirez:2007A&A...465..271R}. $^{(c)}$Differential abundances relative to the Sun using \textsc{isochrones}-based stellar parameters. $^{(d)}$ Weighted average of \ion{Fe}{1} and \ion{Fe}{2}.}
	\begin{tabular}{lccccr} 
		\hline
		\hline
		Element            &   Z   &  TOI-1173 A$^{(a)}$         &  TOI-1173 B$^{(a)}$         &   TOI-1173 A$^{(c)}$         &  TOI-1173 B$^{(c)}$      \\   
		                   &       &     $\Delta$[X/H] (dex)     &  $\Delta$[X/H] (dex)        &    $\Delta$[X/H] (dex)       &  $\Delta$[X/H] (dex)     \\
        \hline   
		\ion{C}{1}         &   6   &    0.100 $\pm$ 0.080        &    0.060 $\pm$ 0.080        &     0.090 $\pm$ 0.070        &    0.050 $\pm$ 0.100     \\
        \ion{N}{1}         &   7   &    0.340 $\pm$ 0.100        &    0.330 $\pm$ 0.100        &     0.380 $\pm$ 0.100        &    0.370 $\pm$ 0.110     \\
        \ion{O}{1}$^{(b)}$ &   8   &    0.185 $\pm$ 0.016        &    0.180 $\pm$ 0.034        &     0.134 $\pm$ 0.041        &    0.180 $\pm$ 0.050     \\
        \ion{Na}{1}        &   11  &    0.137 $\pm$ 0.031        &    0.149 $\pm$ 0.049        &     0.137 $\pm$ 0.041        &    0.139 $\pm$ 0.061     \\
        \ion{Mg}{1}        &   12  &    0.191 $\pm$ 0.016        &    0.153 $\pm$ 0.028        &     0.180 $\pm$ 0.022        &    0.127 $\pm$ 0.029     \\
        \ion{Al}{1}        &   13  &    0.208 $\pm$ 0.022        &    0.182 $\pm$ 0.027        &     0.204 $\pm$ 0.025        &    0.167 $\pm$ 0.033     \\
        \ion{Si}{1}        &   14  &    0.167 $\pm$ 0.012        &    0.125 $\pm$ 0.019        &     0.149 $\pm$ 0.016        &    0.103 $\pm$ 0.025     \\
        \ion{S}{1}         &   16  &    0.110 $\pm$ 0.030        &    0.082 $\pm$ 0.044        &     0.130 $\pm$ 0.050        &    0.070 $\pm$ 0.080     \\
        \ion{K}{1}         &   19  & $-$0.086 $\pm$ 0.029        & $-$0.238 $\pm$ 0.025        &  $-$0.098 $\pm$ 0.029        & $-$0.247 $\pm$ 0.025     \\
        \ion{Ca}{1}        &   20  &    0.112 $\pm$ 0.024        &    0.110 $\pm$ 0.048        &     0.159 $\pm$ 0.057        &    0.135 $\pm$ 0.053     \\
        \ion{Sc}{1}        &   21  &    0.189 $\pm$ 0.051        &    0.200 $\pm$ 0.056        &     0.198 $\pm$ 0.055        &    0.200 $\pm$ 0.048     \\
        \ion{Sc}{2}        &   21  &    0.224 $\pm$ 0.022        &    0.196 $\pm$ 0.047        &     0.166 $\pm$ 0.037        &    0.176 $\pm$ 0.035     \\
        \ion{Ti}{1}        &   22  &    0.170 $\pm$ 0.018        &    0.180 $\pm$ 0.033        &     0.192 $\pm$ 0.043        &    0.142 $\pm$ 0.046     \\
        \ion{Ti}{2}        &   22  &    0.160 $\pm$ 0.036        &    0.150 $\pm$ 0.043        &     0.173 $\pm$ 0.030        &    0.151 $\pm$ 0.036     \\
        \ion{V}{1}         &   23  &    0.184 $\pm$ 0.014        &    0.180 $\pm$ 0.041        &     0.189 $\pm$ 0.045        &    0.171 $\pm$ 0.054     \\
        \ion{Cr}{1}        &   24  &    0.158 $\pm$ 0.012        &    0.105 $\pm$ 0.023        &     0.100 $\pm$ 0.029        &    0.060 $\pm$ 0.035     \\
        \ion{Cr}{2}        &   24  &    0.140 $\pm$ 0.025        &    0.110 $\pm$ 0.080        &     0.110 $\pm$ 0.041        &    0.110 $\pm$ 0.046     \\
        \ion{Mn}{1}        &   25  &    0.108 $\pm$ 0.025        &    0.069 $\pm$ 0.035        &     0.086 $\pm$ 0.041        &    0.109 $\pm$ 0.080     \\
        Fe$^{(d)}$         &   26  &    0.126 $\pm$ 0.010        &    0.107 $\pm$ 0.020        &     0.139 $\pm$ 0.065        &    0.114 $\pm$ 0.069     \\
        \ion{Co}{1}        &   27  &    0.164 $\pm$ 0.014        &    0.112 $\pm$ 0.026        &     0.173 $\pm$ 0.025        &    0.124 $\pm$ 0.025     \\
        \ion{Ni}{1}        &   28  &    0.129 $\pm$ 0.020        &    0.098 $\pm$ 0.023        &     0.081 $\pm$ 0.020        &    0.061 $\pm$ 0.035     \\
        \ion{Cu}{1}        &   29  &    0.202 $\pm$ 0.020        &    0.209 $\pm$ 0.062        &     0.162 $\pm$ 0.031        &    0.182 $\pm$ 0.054     \\
        \ion{Zn}{1}        &   30  &    0.194 $\pm$ 0.029        &    0.206 $\pm$ 0.025        &     0.149 $\pm$ 0.024        &    0.180 $\pm$ 0.060     \\
        \ion{Rb}{1}        &   37  &    0.110 $\pm$ 0.045        &    0.100 $\pm$ 0.045        &     0.110 $\pm$ 0.045        &    0.090 $\pm$ 0.035     \\
		\ion{Y}{2}         &   39  &    0.081 $\pm$ 0.035        & $-$0.006 $\pm$ 0.040        &  $-$0.030 $\pm$ 0.024        & $-$0.125 $\pm$ 0.027     \\
        \ion{Ba}{2}        &   56  &    0.092 $\pm$ 0.025        &    0.040 $\pm$ 0.035        &  $-$0.022 $\pm$ 0.044        & $-$0.113 $\pm$ 0.054     \\
        \ion{Ce}{2}        &   58  &    0.142 $\pm$ 0.027        &    0.020 $\pm$ 0.019        &     0.135 $\pm$ 0.027        &    0.022 $\pm$ 0.020     \\
        \ion{Nd}{2}        &   60  &    0.230 $\pm$ 0.022        &    0.160 $\pm$ 0.042        &     0.217 $\pm$ 0.027        &    0.157 $\pm$ 0.032     \\
        \ion{Eu}{2}        &   63  &    0.208 $\pm$ 0.033        &    0.221 $\pm$ 0.020        &     0.213 $\pm$ 0.033        &    0.250 $\pm$ 0.024     \\
		\hline
		\hline
	\end{tabular}
	\label{tab:chemical_abundances_sun}
\end{table*}

\bibliography{references}{}
\bibliographystyle{aasjournal}

\end{document}